# BUDGET FORECASTING AND INTEGRATED STRATEGIC PLANNING FOR LEADERS

## A Dissertation By

**MATT (MEHDI) SALEHI**
**ORCID iD: 0009-0007-9636-7534**

**California State University, Fullerton**
**Fall, 2025**

---

**In partial fulfillment of the degree:**
Doctor of Education in Community College Leadership

**Department:**
Department of Educational Leadership

**Approval Committee:**
Leobardo Barrera, College of Education, Committee Chair
Nancy Watkins, College of Education
Inez Moore, College of Education, Expert Practitioner

**DOI:**
10.5281/zenodo.15770183

**Abstract:**
This study explored how advanced budgeting techniques and economic indicators influence funding levels and strategic alignment in California Community Colleges (CCCs). Despite widespread implementation of budgeting reforms, many CCCs continue to face challenges aligning financial planning with institutional missions, particularly in supporting diversity, equity, and inclusion (DEI) initiatives. The study used a quantitative correlational design, analyzing 30 years of publicly available economic data, including unemployment rates, GDP growth, and CPI, in relation to CCC funding trends. Results revealed a strong positive correlation between GDP growth and CCC funding levels, as well as between CPI and funding levels, underscoring the predictive value of macroeconomic indicators in budget planning. These findings emphasize the need for educational leaders to integrate economic forecasting into budget planning processes to safeguard institutional effectiveness and sustain programs serving underrepresented student populations.

# TABLE OF CONTENTS

# LIST OF FIGURES

# CHAPTER 1

# INTRODUCTION

Envision a bright-eyed student stepping onto a college campus for the first time, carrying the hopes and dreams of their entire family on their shoulders. They are the first to break the chains and be the first in their family to pursue higher education. But then, the crushing moment comes when they learn that an essential class is unavailable due to funding cuts. In these pivotal moments, the gravity of financial management in higher education can be fully understood. Every dollar in the budget serves as a building block for these dreams, and administrators must navigate the financial seas with a keen eye and steady hand.

This dissertation examined advanced budgeting techniques in California Community Colleges (CCCs) and their alignment with institutional strategic goals and compliance standards among the complexities of financial management in higher education. This study undertook critical analysis to identify the correlation between economic indicators, such as unemployment rates and gross domestic product (GDP) growth, and funding levels for these colleges. Central to this inquiry was integrated planning, where strategic planning guides the college's operational planning and informs the allocation and reallocation of resources, ensuring budgeting practices are aligned with the institution's mission (Salem et al., 2020). It has contributed to the broader discourse on enhancing the efficiency and effectiveness of financial strategies in community colleges.

This chapter introduces the study, delivers the background, clarifies the motivations for tackling this research, and articulates the problem statement. This chapter explains the purpose of the research, with a discussion of its significance to college leaders and educational advocates. Moreover, it defines terms, assumptions, limitations, and delimitations.

## Background of the Problem

The ability to forecast, build, manage, and oversee a budget is an essential competency for all administrators in higher education (Barr & McClellan, 2017). This competency applies to more than just a specific role, as nearly all administrative positions, ranging from entry-level professionals managing minor program budgets to higher positions (e.g., program directors, deans, vice presidents) entail some degree of budget management responsibility (Barr & McClellan, 2017). A comprehensive understanding of the budget and the skills to address budget-related challenges are fundamental for achieving success in higher education (Rabovsky, 2012). Budget management in higher education is a continuous and multifaceted process (Barr & McClellan, 2017). Administrators juggle multiple budgets, including the current operating budget, capital budgets for new facilities, and often budgets for facility management. Addressing these budgetary responsibilities requires involves monitoring expenditures, projecting expenses, and estimating revenues for future fiscal years to support the institution's overall financial plan (Goldberg & Prottas, 2017).

**Management of the Budget Cycle**

Effectively navigating the complex dynamics of financial management requires a clear understanding of the foundational aspects of budget cycle management in higher education institutions. This awareness allows colleges to foresee and respond to financial challenges proactively, ensuring sustainable operations and alignment with their educational mission (Barr & McClellan, 2018). Budget management in higher education is an ongoing process involving multiple budgets simultaneously (Bergmann et al., 2020; Pounder, 1999). The ongoing process of budget management includes monitoring expenditures under the current operating budget, closing the previous year's operating budget, making financial projections for the upcoming fiscal year, managing capital budgets for facilities, and estimating revenues from various sources (Barr & McClellan, 2017). Forecasting, in financial management, generally refers to estimating or predicting how financial variables (e.g., revenues, expenses) will change in the future. Forecasting in the context of budget management in higher education involves making financial projections for future fiscal years, including revenues and expenses (Barr & McClellan, 2017).

Forecasting is not just an administrative task; it is also a requirement highlighted by the Accrediting Commission for Community and Junior Colleges (ACCJC). Forecasting involves the institution engaging in ongoing, meaningful, and collaborative discussions on student outcomes, equity, academic quality, institutional effectiveness, and the continuous enhancement of student learning and achievement (Colleges, 2024). This requirement emphasizes the significance of forward-looking financial planning as a fundamental component of achieving and maintaining accreditation standards, highlighting the critical interaction between financial forecasting and strategic institutional compliance. Such standards urge institutions to incorporate these steps into their budgeting and financial planning processes. The colleges can align their financial resources with their strategic goals and objectives (Hovey, 2012).

The budget cycle also involves developing and implementing the operating budget, which includes creating new budget requests, forecasting revenues and expenses, and managing the processes and procedures for ongoing budget management (Barr & McClellan, 2017). Standard III of ACCJC (2024) defined resource management requirements to assist the institutions' effectiveness in the uses of human, physical, technology, and financial resources to achieve its mission and to improve academic quality and institutional effectiveness. Effectiveness in resource management demands the optimal allocation and use of resources (e.g., finances, human resources, materials, time) in alignment with an organization's objectives. It involves achieving desired outcomes productively, adapting to changes, ensuring sustainability, and making informed decisions (Salem et al., 2020). In this resource management framework, it is essential to focus on fiscal responsibility and stability. This shift underscores the vital connection between strategic resource allocation and educational institutions' financial health and sustainability, setting the stage for a deeper exploration of financial management in higher education (Bers & Head, 2014).

**Fiscal Responsibility and Stability**

Fiscal responsibility and stability refer to the prudent management of financial resources to ensure an organization's or government's long-term financial health and involve setting and adhering to budgets, avoiding excessive debts, ensuring adequate reserves, and making sustainable decisions over time (Laderman et al., 2023). Fiscal responsibility requires transparent accounting practices, informed decision making, and alignment of expenditures with revenues (Barr & McClellan, 2018). One can achieve stability with a balanced budget, low debt levels, and the ability to absorb financial shocks. Fiscal responsibility and stability contribute to the trustworthiness and sustainability of an organization or a government's financial position (Dimitrijevska-Markoski et al., 2021).

Forecasting, moreover, plays a fundamental role in fiscal stability. In revenue and expenditure forecasting within higher education, unit budget managers must collaborate with the institutional budget office to make projections for the current budget year and beyond (Hovey, 2012). Forecasting is inexact, so the estimate of revenues should be conservative, whereas the expenses forecast should be realistic (Hovey, 2012). Historical data, environmental scans, and the unit's specific challenges are keys to forecasting (Goldberg & Prottas, 2017). Additionally, assumptions underlying the forecasts must be explicit to prevent miscommunication. Creating multiple "what if" scenarios to elucidate assumptions and highlight potential issues is beneficial (Hovey, 2012). For instance, projections regarding a new academic program should explain the basis of the projected figures and make assumptions clear, especially concerning funding sources. To ensure financial stability within educational institutions, ACCJC (2024) has defined the following standard that "institutional planning reflects a realistic assessment of financial resource availability, development of financial resources, partnerships, and expenditure requirements" (p. 12).

Understanding how these financial commitments are vital in shaping institutional stability and resource management is essential (Hovey, 2012). This perspective is particularly relevant in the context of ACCJC (2024) standards, which emphasize integrity and purposeful use of financial resources, bridging the gap between theoretical financial practices and their practical application. In the context of higher education institutions' financial management, liabilities refer to the financial obligations or debts that the institution owes to external entities. These can include various obligations such as accounts payable, accrued expenses, deferred revenue, notes and bonds payable, employee benefits, and capital leases (Barr & McClellan, 2018). ACCJC standard III. D. 14 stated, "All financial resources, including short- and long-term debt instruments (e.g., bonds and certificates of participation), auxiliary activities, fund-raising efforts, and grants, are used with integrity in a manner consistent with the intended purpose of the funding source" (p. 24). This transition to liabilities reflects a deeper examination of the financial commitments that institutions must navigate to sustain their operations and strategic initiatives in alignment with ACCJC standards. By effectively managing these liabilities, higher education institutions can maintain financial stability and allocate resources efficiently to improve their cost performance and fulfill their educational missions (Bers & Head, 2014; Evi & Bambang, 2021).

The critical link between strategic financial management, including the handling of liabilities and the alignment with regulatory standards like those of the ACCJC, sets the stage for addressing the core issue of ensuring budgeting techniques not only meet fiscal responsibilities but also reinforce the mission and goals of community colleges, thereby enhancing educational quality and student success (Hovey, 2012). When managing finances in higher education, especially in CCCs, using advanced budgeting techniques allows achieving institutional goals and mandates. Although these techniques provide a sophisticated method for resource management, the difficulty frequently lies in ensuring their compatibility with the institution's primary objective and the principles of effective budget allocation (Hovey, 2012; Li, 2022). To effectively align financial strategy with educational objectives, one must possess refined comprehension and implement budgeting techniques strategically. Nevertheless, the discrepancy in implementing these sophisticated budgeting methods among various community colleges has revealed a distinction between theoretical efficiency and practical efficacy (Barr & McClellan, 2017). This disparity highlights a notable issue in financial management procedures and suggests wider consequences for institutional performance and educational quality.

The core of this problem lies in the potential misalignment of budgeting practices with the institution's mission, leading to a cascade of unfavorable outcomes that can affect every facet of the educational experience offered by community colleges (Hovey, 2012; Sago-Hart, 2022). Ineffective resource use, originating from this misalignment, threatens to deteriorate the quality of education, diminish student success rates, and compromise the ability of these institutions to serve their diverse student bodies effectively (Barr & McClellan, 2017). Community colleges play a vital role in offering higher education opportunities, particularly for underrepresented and nontraditional students. (A. Cohen & Brawer, 2003). Therefore, it is of great importance to address this issue. The ability of community colleges to align their financial strategies with their primary goals is fundamental for maintaining the integrity of the educational mission, optimizing resource allocation, and ensuring the overall success of the student population (Dougherty & Townsend, 2006). This context established the foundation for the subsequent investigation, emphasizing the pressing necessity to evaluate how advanced budgeting techniques align with the goals and effective allocation of funds in CCC. This evaluation is essential to ensuring these vital institutions can continue to fulfill their role in the higher education system efficiently.

## Problem Statement

In the context of community colleges, budgeting plays a pivotal role in shaping the institution's ability to meet its mission of providing accessible, quality education. The adoption of advanced budgeting techniques in community colleges has become increasingly prevalent in recent years as institutions strive to optimize resource allocation and maintain financial sustainability (Salem et al., 2020). However, these budgeting practices also influence the support and expansion of programs central to advancing diversity, equity, and inclusion (DEI), such as DEI Offices, Black Scholar Unions, disability centers, extended opportunity programs and services (EOPS), and TRIO programs. These initiatives are vital for reducing systemic inequities and fostering a supportive educational environment, particularly for underserved and underrepresented student populations (Reckhow et al., 2021). Although these programs have demonstrated potential, their success often depends on consistent and reliable funding, which can be affected by institutional leaders' uncertainties regarding the continuity of financial resources. Several scholars have argued the successful implementation of advance budgeting techniques requires alignment with the institution's strategic mission and educational goals (Barr & McClellan, 2017; Hovey, 2012). When there is a misalignment between budgeting practices and institutional priorities, resources may be allocated inefficiently, undermining the college's ability to provide high-quality education (Salem et al., 2020). As institutions face growing financial pressures, the need for efficiency and strategic use of resources becomes highlighted to sustain student success, particularly for those from underrepresented groups and minorities. Yet, evidence has suggested many community colleges struggle to achieve this alignment, particularly in the face of shifting economic conditions (Ely & Calabrese, 2016). The problem this study addressed was the concern that arises when the use of advanced budgeting techniques in community colleges does not align with the institution's mission and efficient budget allocation. These misalignments may lead to ineffective resource use, diminishing the quality of education and potentially compromising student success.

## Purpose Statement

The primary objective of this research was to evaluate how well-advanced budgeting techniques, including economic indicators, meet the standards for fiscal responsibility and resource management in CCCs. To do this, the study investigated relationships between economic variables and funding levels, as well as the predictive value of these variables for future budget decisions. The study is guided by the following research questions, which aim to examine the connections between advanced budgeting techniques, fiscal policies, and funding levels for CCCs and their students. These questions highlight the focus areas of the research.

**Research Questions**

1.  What is the relationship between the advanced budgeting techniques' economic indicators (e.g., unemployment rates, GDP growth) and funding levels for California Community Colleges and students?

2.  What is the relationship between fiscal policies (i.e., policies impacting income, corporation, sales, and local property taxes) and funding levels for California Community Colleges and students?

**Significance of Study**

This research was important and made make a significant contribution to educational leadership because it has highlighted the critical role that effective budget management and precise forecasting play in protecting CCCs' sustainability and efficacy. D. Drury (1978), in his historic book, stated CCCs advocate for accessible, affordable, and inclusive education, which depends considerably on well-founded financial strategies. Budget management includes prioritizing basic aspects of students' educational experiences and allocating resources effectively (Salem et al., 2020). Budget forecasting is an essential instrument that assists colleges in estimating their revenue flows and expenditures (Hovey, 2012). It fosters knowledgeable and data-centered decision making for strategic management. Sound budget planning is fundamental to colleges' mission as it establishes the foundation for developing and maintaining fundamental programs and services, ensuring students receive the necessary tools for success (Goldberg & Prottas, 2017).

The study's findings were significant for advancing social justice and DEI initiatives in CCCs. Many programs that address systemic inequities, such as DEI Offices, Black Scholar Unions, disability resource centers, EOPS, and TRIO, depend heavily on strategic budgeting and reliable funding (California Community College Chancellor's Office [CCCCO], 2023). These initiatives assist underserved and underrepresented students; however, budget uncertainties and misalloction of resources often limit their effectiveness (Brooks, 2019). By examining how advanced budgeting techniques align with institutional missions, this study provides insights that could enhance the financial sustainability of these programs, ensuring they continue to promote equity and foster inclusive learning environments. Understanding how economic variables influence budgetary decisions empowers educational leaders to allocate resources strategically (Califorornia Community College Chancellor's Office [CCCCO], 2022), directly addressing the needs of marginalized communities.

Furthermore, this study's significance escalates when considering the potential risks economic crises can bring to the financial stability of community colleges. During economic recessions, like the COVID-19 global pandemic economic crisis, there is a likelihood for colleges to encounter severe funding and resource cutbacks (Barr & McClellan, 2018; Rosinger et al., 2022). More preparation is needed to ensure colleges' ability to accomplish their missions. However, adopting well-built budget management and identifying forecasting enables colleges to formulate contingency strategies and effectively distribute resources in challenging times (Hovey, 2012; Li, 2022). By exercising fiscal responsibility and taking a forward-thinking perspective, colleges can maintain resilience and continue delivering necessary educational services to their communities, even though economic hardships (Jimenez, 2014; Rosinger et al., 2022).

This study has also delivered invaluable wisdom for various stakeholders in the educational system, including college presidents, superintendents, administrators, faculty, instructors, student affairs professionals, students, board members, policymakers, parents, and community leaders. For educational leaders and administrators, the findings can guide more effective and strategic financial planning, aligning budget management practices with institutional goals and assuring the sustainability of quality educational programs. Faculty and instructors may find the outcomes beneficial for understanding the financial support that sustains their teaching and research activities. Students and parents can gain reassurance from the institution's commitment to fiscal responsibility, which directly impacts the quality and accessibility of education. Finally, board members and policymakers can use the study's insights for informed decision making.

**Scope of the Study**

This study meticulously examined the relationship between economic indicators and funding levels in CCCs. It optimized budgeting practices to align with strategic goals and ensure fiscal responsibility. It discussed the nuanced correlation between variables such as unemployment rates, GDP growth, and the allocation of financial resources to these educational institutions. However, it is essential to note the scope of this research was limited to economic indicators and their direct impact on funding mechanisms, excluding broader sociopolitical factors that may also influence budgeting decisions. Although recognizing the complex system in which community colleges operate, this investigation did not cover the detailed analysis of one specific college management practice or the specific educational outcomes resulting from fiscal strategies. The study aimed to provide a macrolevel understanding of financial planning in ongoing economic conditions. It has offered insights that can inform broader budgetary adjustments and policy formulations.

**Assumptions of the Study**

This study's economic indicators (e.g., unemployment rates, GDP growth) effectively represented CCCs' economic conditions. These indicators were critical for assessing funding levels and dependable measures of the overall economic climate. This assumption was based on the belief that economic patterns directly influence state budget allocations (Popesko et al., 2016); therefore, the financing is accessible to community colleges. The association between these variables and college funding relies on the accuracy and significance of these economic data (Hovey, 2012).

The other assumption was that the budgeting methods used by CCCs are to align with the institutions' strategic objectives and regulatory requirements. This assumes colleges are willing and able to employ sophisticated budgeting procedures properly. This study acknowledged these schools prioritize economic cautiousness and resource management to maintain their educational goal. This assumption recognized institutions are dedicated to maximizing their financial resources to improve the quality of education and student achievement.

Lastly, this study assumed the data collected from CCCs, including funding levels, budgeting practices, and strategic planning documents, were accurate and reflected the current practices and conditions in these institutions. This assumption was essential for analyzing how effectively advanced budgeting techniques adhere to standards for fiscal responsibility and resource management. The integrity and completeness of these data were critical for understanding the relationship between economic variables and funding levels, as well as predicting future budgetary adjustments (Hovey, 2012). The study was based on these assumptions, which enable a detailed analysis of how economic indicators and community college financing in California interact.

**Study Delimitations**

This study deliberately delimited its focus to CCCs, excluding other higher education institutions (e.g., public and private universities). This delimitation was based on two reasons. First, community colleges serve a unique role within the broader landscape of higher education, often catering to nontraditional students, providing vocational training, and offering accessible education to a diverse demographic (Pasadena City College, 2002). This unique aim affects how they receive financing and manage their budgets, giving them a particular and essential subject for studying how economic indicators affect budgeting methods (Barr & McClellan, 2018). Second, by concentrating on community colleges, the study could dive deeper into the specific challenges and opportunities these institutions face, offering more targeted understandings that might be less relevant in the context of universities. This concentration strengthened the study by offering specialized information and practical recommendations customized to the distinct context of community colleges.

This study was restricted geographically to California. California's varied economy and substantial size made it an exceptional subject for studying how economic variables relate to funding for community colleges. This geographical restriction allowed for a more detailed analysis of state-specific policies, economic conditions, and funding frameworks directly impacting community colleges. This constraint focused on studying community colleges only in California, providing a detailed understanding of the state's unique context. This focus on studying community colleges in California could be a valuable case study for future comparative analysis.

Furthermore, the study restricted its analysis to specific economic indicators, namely unemployment rates and GDP growth, to assess their correlation with funding levels for community colleges. The study analyzed how broader economic trends can affect institutions' financial health and budgeting decisions by concentrating on specific metrics. I chose this delimitation with the understanding other economic variables may influence community college funding, concentrating on unemployment rates and GDP growth allowed for a focused investigation of the most directly relevant and widely recognized indicators. This method enhanced the research by guaranteeing clarity and significance in its results and recognized a more comprehensive collection of indicators could provide further insights.

**Study Limitations**

One of the primary limitations of this study was the reliance on publicly available data concerning the funding levels and budgeting practices of CCCs. These data offered a basic understanding of financial trends and practices but may have only partially encompassed specific institutions' intricate budget management techniques or sophisticated decision-making processes. The inherent limitation of secondary data means that specific details, particularly regarding the internal deliberations or rationale behind specific budgeting choices, remain beyond the scope of this analysis. This limitation hindered the study from thoroughly examining the complexities of financial planning and administration in these universities.

A noteworthy limitation of this study stemmed from its reliance on historical data spanning the past 30 years to analyze funding levels and budgeting practices in CCCs. Although historical data help recognize patterns and make connections, it may not reliably forecast future financial situations or the efficiency of budgeting methods in a quickly evolving economic setting. This temporal disconnection challenges forecasting future budget increases or reductions with precision, as past trends may not fully account for unforeseen economic fluctuations or shifts in educational funding policies.

Additionally, this study's focus on economic indicators (e.g., unemployment rates, GDP growth) as predictors of funding levels may have only accounted for some factors influencing budget allocations to community colleges. Additional factors, including political decisions, policy changes, and demographic fluctuations, are essential in influencing funding levels but were not the focus of this study. Although these factors were acknowledged, the study needed to thoroughly examine their direct impact, which presented a limitation in understanding the complete picture of what influences community college funding in California.

The study's design and depth limitations indicated the need for additional research to fill these gaps. Although the study's design and depth limitations presented challenges to the completeness of the analysis, they also identified topics for further investigation that could enhance the results of this study and provide a more complete perspective on financial strategic planning in community colleges.

## Definitions of Key Terms

The following terms are defined as important for this study:

*Budget*. Budget is a detailed list of projected expenses for a specific timeframe, including suggestions on how to fund them (The American Heritage Dictionary of the English Language, 2016).

*Budgeting*. Budgeting is the act of preplanning and determining the expected expenses for a specific period, typically in a precise manner (The American Heritage Dictionary of the English Language, 2016).

*Contingency planning*. Contingency planning is to predict any external or internal obstacles and create alternative methods in case they happen (Friedman, 2013a).

*External funding*. External funding pertains to financial sources that are not part of the regular federal, state, and local funds received annually. The sources of funding encompass grants, donations, endowments, and scholarships as stated by Hill (2003).

*Federal funds rate.* The federal funds rate is the interest rate that banks with surplus reserves charge other banks for short-term loans. Therefore, it frequently serves as a primary indication of the general national interest rates (Mishkin, 2019).

*Financing*. Financing refers to the provision of finances or capital (Stapleton, 1981).

*Forecasting*. Forecasting is a strategic technique that aids management in dealing with future uncertainties. The process begins with specific assumptions derived from the management's experience, knowledge, and judgment. These projections are based on quantitative methodologies and are used to forecast future months or years (Friedman, 2013b).

*General obligation bond.* The general obligation bond is a municipal bond secured by the revenue-generating authority of the issuer, typically through property tax collections. The Standard & Poor's general obligation bond rating for the state of California assesses the state's overall creditworthiness and its future capacity to repay a general obligation bond. An AAA rating is the highest, followed by AA, A, BBB, and so forth. Ratings are adjusted by employing plus (+) or negative (-) indications (Hardy & Runnels, 2014).

*Gross domestic product.* The gross domestic product (GDP) is the sum of the monetary value of all finished goods and services produced within the country's borders during a set timeframe. GDP is a comprehensive metric that reflects a country's total economic output and is frequently used to assess a nation's economic well-being and living standards. Mankiw (2020) defined GDP as the total of consumer consumption expenditures, business investments, government spending, and net exports (i.e., exports minus imports).

*Integrated planning/budgeting*. The integrated planning/budgeting operates on the principle that planning influences resource allocation choices (i.e., budgeting). The college's vision and mission guide resource allocations to enhance college instructional programs, departments, services, and student activities for continual improvement (Colleges, 2024).

*Leading indicator variable.* The leading indicator variable is a variable that typically anticipates or precedes the behavior of the entire economy. It will decrease before the rest of the economy collapses and increase before a peak emerges in the rest of the economy (Morris & Morris, 2007).

*Strategic planning.* Strategic planning is a methodical process where an organization establishes and secures support from important stakeholders for priorities essential to its mission and adaptable to the surroundings. Strategic planning directs the procurement and distribution of resources to accomplish these aims (Allison & Kaye, 2005).

*Trailing indicator variable.* The training indicator variable is one that typically follows the behavior of the wider economy. It will decrease once the economy declines and increases after the economy reaches its peak. It is not a valid predictor variable; instead, it confirms previous changes in activity (Morris & Morris, 2007).

*Transformative leadership.* Transformative leadership refers to a leadership style where the leader recognizes necessary changes, establishes a vision to steer the change through inspiration, and implements the change with the group members' dedication (Friedman, 2013c).

## Organization of the Dissertation

Chapter 1 presented a detailed summary of the financial management difficulties encountered by CCCs, outlining the problem and goal of this study. I looked further into the need for efficient budget management and forecasting to sustain and enhance the effectiveness of these institutions. I outlined the scope of the study, along with assumptions, delimitations, and limitations, to clarify the extent and boundaries of the research. I clearly explained key terms necessary for comprehending the study's context.

Chapter 2 provides an in-depth review of the research concerning strategic planning, decision making, budgeting methods, economic indicators, and how they influence the financing levels of community colleges. This review lays down the theoretical groundwork for the investigation and identifies deficiencies in the current research. Chapter 3 describes the research design and methodology, detailing the data collection and analysis methods employed to investigate the correlation between economic indicators and funding levels for CCCs. Chapter 4 thoroughly examines the study's results, including an in-depth analysis of the gathered data and its significance in comprehending the correlation between economic variables and budgeting procedures in community colleges. In Chapter 5, I present the conclusions derived from the study's findings and their interpretations. These recommendations aim to enhance budgeting strategies in community colleges to better align with their strategic goals and compliance standards.

# CHAPTER 2

# REVIEW OF THE LITERATURE

This chapter provides a comprehensive review of the literature that supports this study. It begins by exploring the study's foundations, including historical, philosophical, and theoretical perspectives that frame the research. Next, it examines the scholarly empirical literature to contextualize the economic indicators and fiscal policies affecting California Community Colleges (CCCs). Finally, the chapter presents the study's conceptual foundation, synthesizing key concepts and relationships relevant to the research. The chapter concludes with a summary, highlighting the critical insights gained from the review.

## Historical, Philosophical, and Theoretical Foundations

The historical foundation of this study, discussed in this section, involves the evolution of strategic planning and budgeting in U.S. businesses and academic institutions, specifically CCCs. This section will further explore the philosophical foundations of these techniques, primarily focusing on pragmatism. Additionally, this section will lay out the theoretical framework of the study, examining resource dependence theory (RDT) and new public management (NPM). Understanding these elements provides an essential context and frame for investigating the correlation between economic indicators and funding levels in CCCs. These foundations help answer the study's core research question. This comprehensive approach helps understand the current state of fiscal responsibility and resource management in CCCs.

### Historical Foundation

The historical section in this dissertation explores the development of strategic planning in business and academic domains. This discussion aims to uncover the important events that have shaped contemporary strategic planning.

*Businesses*

Strategic planning has evolved significantly, tracing its roots to military and business applications. U.S. business and industry had a global comparative advantage in the post-World War II era. Growth was the primary goal, and planning focused on long-term objectives, with firms often projecting business actions outward for as much as 20 years (Drucker, 2012; Hovey, 2012). In the boom years following World War II, formalized planning was used primarily in business and industry for project planning and growth management. However, the Vietnam War and the subsequent environmental change forced businesses to shift their planning focus from long- to shorter-term objectives. This shift was necessary to ensure businesses could adapt to the rapidly changing economic and political landscape (Mintzberg, 1994).

Several factors led to the shift in strategic planning from pursuing long-term objectives to a shorter planning horizon. One of the most important factors was the Vietnam War, which led to economic instability and uncertainty. Another factor was the increasing pace of technological change. New technologies were emerging rapidly, and businesses needed to be able to adapt quickly to these changes. Long-term planning became less effective in this environment, as businesses could not accurately predict how future technologies would impact their industries (Mintzberg, 1994).

The shift in strategic planning to a shorter horizon presents several business challenges and opportunities. One of the biggest challenges is the need to balance short- and long-term objectives, as businesses need to succeed in the long term. The next challenge is the need to be adaptable. The business environment is constantly changing, and businesses need to be able to adapt quickly to these changes.

Despite the challenges, the shift in strategic planning to a shorter horizon also presents several business opportunities. One of the most significant opportunities is the ability to be more responsive to customer needs. Businesses can now quickly respond to changes in customer preferences and demands, which can help them stay ahead of the competition. Another opportunity is the ability to experiment with new technologies and business models (Onabanjo A, 2024; Teece, 2007).

### Academic Institutions

Similarly, academic institutions like CCCs have also integrated strategic planning into their operations, albeit at a slower pace. Academic institutions have traditionally been slow to design and implement planning processes as effective (Allison & Kaye, 2005; Bryson, 2017). The general adoption of planning processes in academic institutions have reflected the route of development in business and industry, but with a lag in the adaptation to strategic planning processes. Strategic planning models gained widespread prominence in higher education in the 1980s (Bryson, 2017).

One reason for the slow adoption of strategic planning in academic institutions is the unique nature of their environment. Academic institutions are complex organizations with various stakeholders, including students, faculty, staff, alumni, and donors. These stakeholders often have different and sometimes conflicting priorities. Another reason for the slow adoption of strategic planning is that various sources, including government funding, tuition and fees, and private donations, typically fund academic institutions (Rowley et al., 2002).

Given this historical context and the unique challenges that academic institutions face, exploring the philosophical reinforcements that drive strategic planning in CCCs becomes compulsory. This inquiry helps better explain the complexities of aligning advanced budgeting techniques with institutional goals and compliance standards.

### California Community Colleges

As a part of the broader American community college system, CCCs has a rich history that dates to the early 20th century (Pasadena City College, 2002). This system emerged as a response to multiple social forces, including the growing need for a workforce adept at managing the country's flourishing industries and a quest for social equality through enhanced access to higher education (A. Cohen & Brawer, 2003).

A vital component that fueled the development of community colleges was the recognition of science and education as catalysts for societal progress (D. Drury, 1978). Schools started to expand their programs to accommodate the diverse goals of an increasing population. As a result, community colleges, including those in California, grew rapidly (Pasadena City College, 2002). Furthermore, like its counterparts nationwide, the California Community College system historically began to shoulder enormous societal responsibilities (D. Drury, 1978). These colleges aimed to address and provide solutions to social and personal challenges ranging from racial integration to unemployment (D. Drury, 1978). In essence, society placed faith in educational institutions to ease societal issues.

In the 21st century, the mission of the CCCs is rooted in these historical developments (Brooks, 2019). The system seeks to provide accessible, high-quality education that equips students with the knowledge and skills necessary for success in an increasingly complex world. It aims to serve diverse communities, contribute to workforce development, and address societal challenges through education (A. Cohen & Brawer, 2003).

As CCCs quickly grew into more complex institutions, they naturally underwent significant transformations in their financial management practices (Dougherty & Natow, 2015). One must recognize the integral role financial management plays in ensuring these institutions continue to fulfill their mission (Bakhit, 2014). The trends in financing community colleges have evolved with institutional objectives and structure shifts over time. For instance, some colleges manage budgets exceeding $100 million (Ely & Calabrese, 2016).

Community colleges, which began primarily as extensions of secondary schools, shifted their financial ground as independent community college districts (Dougherty & Townsend, 2006). Historically, community colleges had to adapt to different sources of funds, including local taxes, state revenues, and tuition (A. Cohen & Brawer, 2003). To provide a comprehensive understanding of the evolving financial landscape, reviewing the historical context of specific laws and regulations that have significantly influenced their budgeting practices is critical. This historical ground clarifies some complexities and challenges these institutions face in aligning their budgeting strategies with their mission and goals, highlighting the importance of integrated planning.

Some laws and regulations impacted CCCs' budgeting, including Proposition 13, passed in 1978; Proposition 98, passed in 1988; Assembly Bill 1200, enacted in 1991; and Proposition 25, passed in 2010 (Barr & McClellan, 2018). California's Proposition 13, passed in the late 1970s, is a case in point; it limited property taxes and compelled community colleges to turn to the state for funding. Consequently, within two years, the state's share of community college revenues skyrocketed from 42% to nearly 80%. Proposition 98 significantly impacted the funding of California's public schools and community colleges (Vasquez Heilig et al., 2014). Proposition 98 established a minimum funding requirement for K–14 education, including kindergarten through 12th grade and community colleges. It has significantly impacted community college funding in California, providing minimum funding; however, approximately only 11% of Proposition 98 funding goes to community colleges, with the remaining funds going to K–12 schools (Reckhow et al., 2021). One of the most effective laws impacting the college system is Assembly Bill (AB) 1200. Passed in 1991, it sought to address fiscal accountability within California school districts to avoid insolvency and the subsequent state financial bailouts. However, the focus of AB 1200 was mainly on K–12 school districts rather than directly on community colleges. AB 1200 created a more robust framework for fiscal accountability in education in California by establishing the Fiscal Crisis and Management Assistance Team (Barr & McClellan, 2018). California's Proposition 25, passed in 2010, had significant implications for the state's budget process, indirectly affecting funding for community colleges. The proposition lowered the legislative vote requirement to pass a budget from two thirds to a simple majority. However, it did not alter the two thirds vote requirement for raising taxes. It expedites budget approval which can result in more predictable and timely funding, allowing community colleges to plan and allocate resources more efficiently (Sinclair, 2017).

Additionally, expenditure patterns have experienced fluctuations with spending per student, which indicates various trends over the years. These fluctuations have been affected by several factors, including changes in student enrollment patterns and economic downturns such as the Great Recession of 2008–2011 (Jimenez, 2014) and the COVID-19 global pandemic in 2020. Community colleges have often resorted to measures (e.g., increasing student fees, deferring maintenance, hiring part-time faculty, freezing other employment) to mitigate shortfalls during lean years (Levin et al., 2006). Financial management in community colleges, including CCCs, is an ever evolving and complex process. Understanding these financial trends and adapting to the changes is vital for these institutions to fulfill their mission and serve diverse communities effectively (Barr & McClellan, 2018).

**Philosophical Foundation**

Pragmatism is a philosophy that emphasizes the importance of doing things in the real world to test ideas and see if they work (Dennes, 1940). It was developed in the late 19th century in the United States by thinkers such as Charles Sanders Peirce, William James, and John Dewey (Nelson, 2019). The philosophical approach of pragmatism served as the foundation for this study's emphasis on useful and applicable research findings that directly addressed pressing problems with financial management in CCCs.

Pragmatists believe ideas are only meaningful if they can be implemented and tested in the real world. They also believe that the world is constantly changing, so individuals' ideas should change too. Pragmatists are more interested in finding practical solutions than developing abstract theories (Dennes, 1940; James, 2010). Pragmatism, moreover, allows for a pluralistic approach to research. A pluralistic approach is a way of thinking and acting that embraces diversity and recognizes that there is no single right way to do things (Morgan, 2014). It is based on the belief that different people have different perspectives and experiences and that all of them are valuable (Johnson & Onwuegbuzie, 2004; Morgan, 2014). Pragmatists argue the research question is more important than the method or paradigm underlying the question (Creswell, 2014). This study focused on addressing the misalignment between budgeting techniques and strategic goals as a problem-centered pragmatic approach.

Pragmatism is action oriented and seeks to produce results that have real-world applications (Johnson & Onwuegbuzie, 2004). By understanding correlations and predicting future budgeting decisions, this study has provided actionable insights for CCCs to ensure better fiscal responsibility and resource management. Pragmatism does not seek an ultimate, unchanging truth but rather views truth as what is most useful, practical, and effective in real-world situations (James, 2010). In the context of this study, a pragmatic perspective would not merely be about finding correlations or predictive variables but about determining what works best for effective resource use and compliance in the dynamic educational landscape.

Moreover, pragmatism acknowledges that research is value laden (Morgan, 2014). This study valued effective resource management, fiscal responsibility, and academic quality. Through a pragmatic lens, these values can be integrated and central to the research process. When examining the correlation between fiscal policies and funding levels for CCCs, a pragmatic approach would encourage me to be open to diverse data sources (i.e., both qualitative and quantitative) and consider various economic theories that might illuminate the research problem (Johnson & Onwuegbuzie, 2004). The aim would not just be about determining the strength or direction of the correlation but about understanding what these relationships mean in the broader context of resource management and quality education.

This study investigated the correlations between economic indicators and funding levels and examined the predictive value of these relationships. Pragmatism offers flexibility in integrating diverse methods and theoretical perspectives that cater to this dual purpose (Johnson & Onwuegbuzie, 2004; Morgan, 2014). Besides, this study's pragmatic and pluralistic approach enhanced its methodology by embracing various perspectives and methodologies. This complexity reflected the complex nature of financial management in community colleges. The focus on pragmatism highlighted the significance of practical and implementable research results that can directly guide strategic planning and budgeting procedures (Morgan, 2014). This examination embraced a pluralistic research approach, which involved considering and actively pursuing a range of economic theories, analytical tools, and interpretative frameworks (Johnson & Onwuegbuzie, 2004; Morgan, 2014). This technique facilitated a more thorough comprehension of how economic indices, such as unemployment rates and gross domestic product (GDP) growth, might impact the allocation of funds to community colleges. The statement recognized the complex nature of economic effects on education finance necessitates a comprehensive analytical methodology that can adjust and address the subtle variations across community colleges and their distinct financial situations.

Furthermore, this research's practical basis supported the idea of a dynamic interchange between theory and practice. As it suggested, theories about budget management and strategic planning should be regularly evaluated considering the challenges faced by community colleges (James, 2010; Johnson & Onwuegbuzie, 2004). The results of these evaluations should then be used to inform academic research and practical implementation within the institutions. This iterative process guaranteed the study stayed firmly rooted in pragmatism, offering practical insights that community colleges can use to handle the difficulties of changing economic situations. This study has enhanced the academic conversation on higher education finance by combining various approaches and viewpoints. Additionally, it has provided a practical guide for community colleges to align their budgeting practices with their strategic objectives in a constantly evolving economic landscape.

**Theoretical Framework**

The theoretical framework of this dissertation provided a structured analysis for how external economic indicators influence budgeting practices in CCCs. It integrates RDT and pragmatism to explore the dynamic interaction between these institutions and their economic environment, underscoring the necessity for strategic alignment of budgeting processes with institutional missions and strategic goals.

*RDT*

RDT asserts that organizations depend on external resources to achieve their goals and must interact with and adapt to their external environments to obtain these resources (Hillman et al., 2009). Consequently, these external entities and conditions can influence an organization's behavior and decision-making processes. The theory emphasizes the importance of understanding and managing external relationships and dependencies to ensure the organization's survival and success. As a result, organizations must adapt their strategies and operations to remain competitive and viable (Hillman et al., 2009; Hovey, 2012). In the context of this research study, resource dependence theory helped examine how community colleges use advanced budgeting techniques to secure resources from their stakeholders, such as state and local governments, businesses, and donor organizations.

Scholars have used RDT as a practical tool to perform studies on financial management in public and private organizations. For example, Hillman et al. (2009) reviewed the existing literature on RDT and budgeting. The authors found that RDT has been used to examine various budgeting issues, such as the relationship between budgeting and organizational performance, the impact of environmental uncertainty on budgeting, and the role of power and influence in budgeting decisions. Pfefer and Salancik (1978) explained the impact of resource dependence on budgeting practices in higher education and the impact of resource dependence on budgeting practices. The authors found that colleges and universities more dependent on external resources are more likely to use centralized budgeting processes. The authors also found colleges and universities that are more dependent on external resources are more likely to use budgeting as a tool for control and coordination. Whereas RDT offers valuable insights into how organizations adapt to external conditions to obtain necessary resources, NPM presents another critical framework for understanding how public sector organizations, like community colleges, can optimize performance through managerial practices rooted in efficiency, effectiveness, and accountability.

*NPM*

NPM is a philosophy of public administration that emphasizes the importance of efficiency, effectiveness, and accountability. It is based on the belief the private sector can provide better services than the public sector and that public sector organizations should adopt business practices to improve their performance (Hood, 1991; Osborne & Gaebler, 1993). NPM emerged in the 1980s, and many governments in many countries have adopted it because then (Hood, 1991). It has led to several changes in public sector organizations, including an increased emphasis on performance measurement and accountability, decentralization of decision making, introduction of market-based mechanisms (e.g., contracting out, competition) and a greater focus on customer satisfaction (Pollitt & Bouckaert, 2011).

NPM advocates adopting business practices in the public sector (Diefenbach, 2009). Scholars have widely adopted NPM in higher education in recent years, leading to a greater emphasis on efficiency and accountability in budgeting and other areas of operations. For instance, the colleges have implemented advanced budgeting techniques through a new financial model called the "Activity-Based Costing" system. This model focuses on aligning resources more effectively with the strategic goals of the university by tracking the costs of different activities more rigorously, thus promoting greater accountability and transparency in budget allocation (Tirol-Carmody et al., 2020). NPM examines how advanced budgeting techniques can improve the efficiency and accountability of budgeting in community colleges. For example, the budgeting processes of community colleges that use advanced budgeting techniques can be compared to the budgeting processes of community colleges that do not use advanced budgeting techniques.

RDT and NPM offered valuable frameworks for understanding the complexities of budgeting and resource management in CCCs. RDT helped illuminate how these institutions depend on various external resources, including state funding and local partnerships, to achieve their strategic objectives (Hillman et al., 2009). On the other hand, NPM provided insights into the administrative philosophies that prioritize efficiency and accountability, which are integral to advanced budgeting techniques (Hood, 1991; Pollitt & Bouckaert, 2011). In the context of this study, integrating these theories can provide a comprehensive analysis of how community colleges in California adopt advanced budgeting techniques, such as using economic indicators to not only meet the standards of fiscal responsibility but also to bring effective resource management.

## Empirical Literature Review

The role of budgeting in educational institutions has been an area of sustained academic inquiry, given its direct impact on resource allocation and, ultimately, the quality of education (Barr & McClellan, 2017; Cheslock, 2010). Whereas budgeting practices in 4-year universities have received considerable attention, fewer studies have focused on community colleges, especially those in California (J. Williams, 2000). This importance is mainly considerable because these colleges face unique financial challenges and are essential players in higher education, providing accessible and affordable education to diverse populations (A. Cohen & Brawer, 2003). This literature review focuses on strategic integrated planning and resource allocation processes at community colleges. Integrated planning refers to a planning process aligned with the budgeting process (J. Williams, 2000). In an integrated model, the budget planning and resource allocation processes cannot exist as separate entities from the college-wide strategic planning process (Barr & McClellan, 2017; Salem et al., 2020). The primary objective of integration is to bring the college's resource allocation into greater conformity with the stated mission and collective understanding of college objectives (Salem et al., 2020). This approach yields measurable outcomes, with forecast costs and identified means to cover those costs, like the outcomes of a traditional strategic planning process in business and industry (Salem et al., 2020). However, despite the benefits of integrated planning, CCCs face critical challenges that interfere with their ability to implement these strategies effectively (J. Williams, 2000).

### Critical Challenges in CCCs

The community college landscape in California is burdened with challenges that require robust planning and strategy (Boggs & McPhail, 2016). A veteran community college consultant, interviewed, has pointed out that accreditation issues and weak planning processes are significant challenges facing these educational institutions (Frost, 2009). Similarly, Phelan (2005) emphasized planning consumes the most substantial portion of time for community college presidents, reiterating the importance of effective planning in sustaining financial support for colleges.

Whereas the preceding challenges are internal to the institutions, external factors add another layer of complexity. Environmental changes (e.g., shifts in the political climate, economic instabilities, societal trends) have a direct impact on community colleges (Brooks, 2019). These changes can affect everything from enrollment rates to the availability of state and federal funding, forcing colleges to be agile and responsive to maintain financial viability and educational quality (Booze, 2019; Hovey, 2012).

Perhaps the most critical challenge is environmental volatility, a constantly shifting landscape community colleges must navigate (Bailey, 2002). This volatility can manifest in various ways, such as sudden changes in government funding, local economic downturns, or global pandemics (Phelan, 2005). The dynamic nature of these external factors demands that community colleges not only adapt, but do so quickly, necessitating the use of advanced budgeting techniques to forecast various scenarios and plan accordingly (Goldberg & Prottas, 2017). The need to adapt to environmental volatility makes integrating strategic planning and resource allocation even more critical (J. Williams, 2000).

**Strategic Planning**

Strategic planning is an indispensable tool for educational institutions aiming to align their organizational objectives with resource allocation (Falqueto et al., 2020). Strategic planning is a systematic process of envisioning a desired future and translating this vision into broadly defined goals, objectives, and a sequence of steps to achieve them (Ward, 2018). George (2017) found strategic planning can be effective in public organizations when implemented correctly. The study also found the success of strategic planning depends on various factors, including leadership, stakeholder involvement, and the use of performance indicators. Moreover, Elbanna et al. (2016) examined the role of formal strategic planning and found formal strategic planning has a strong positive relationship with implementation success in public service organizations. Formal strategic planning is a conventional approach that involves an intentional and sequential process of formulating and implementing plans. This approach typically involves environmental scanning, goal setting, strategy formulation, implementation planning, and constant evaluation (Bryson, 2017). Elbanna et al. (2016) suggested organizations that engage in formal strategic planning are more likely to achieve successful implementation of their strategies. However, there are individual-level predictors of commitment to strategic plans among planning team members in public organizations. Cognitive styles and user acceptance significantly predict commitment to strategic plans among planning team members (George et al., 2018). It is imperative to recognize strategic planning does not function in isolation. Instead, it serves as the support for informed and effective strategic decision making, a concept that further emphasizes the role of planning in achieving long-term objectives (Bryson, 2017; George et al., 2016).

Building upon this foundation, the subsequent sections discuss the nuances of strategic planning, exploring its vital role in organizational decision making, the evolution and advancement of planning models, and a detailed analysis of various strategic planning models. These discussions provide a comprehensive understanding of how strategic planning shapes and guides the rotation of organizations in achieving their long-term goals.

### *Strategic Planning and Decision Making in Organizations*

Strategic decision making is making decisions aimed at achieving a long-term goal or objective (George et al., 2018). Strategic decision making involves analyzing the current situation, assessing available resources, and determining the best action to achieve the desired result (Dimitrijevska-Markoski et al., 2021). It requires careful consideration of all possible options and weighing each option's potential risks and benefits (Falqueto et al., 2020). George et al. (2016) found strategic planning improves strategic decision quality in public organizations by using a systematic approach and involvement of top policymakers, managers, lower-level staff, and external stakeholders. Furthermore, they found the formality of the strategic planning process positively affects strategic decision quality.

Desmidt and Meyfroodt (2021) uncovered those attitudes toward strategic planning, perceived pressure to use strategic plans, and perceived behavioral control are related to politicians' intention and use of strategic plans. They used a theoretical model based on Ajzen's theory of planned behavior. A theoretical model known as the theory of planned behavior explains how people's attitudes, subjective norms, and perceived behavioral control affect their intentions and behaviors (Ajzen & Klobas, 2013). The theory of planned behavior suggests if politicians have positive attitudes toward strategic planning and feel in control of their ability to use it effectively, they are more likely to use it as a decision-making tool (Tornikoski & Maalaoui, 2019). This finding helps in motivating higher education leaders to integrate strategic planning and resource allocation processes. After exploring the interplay between strategic planning and strategic decision making, the evolution of planning models becomes fundamental to examine how planning models themselves have advanced to meet the complex challenges of resource allocation and strategic alignment in educational institutions (De Lancer Julnes et al., 2020; Desmidt & Meyfroodt, 2021). Recent literature has highlighted a noticeable trend toward adopting advanced planning models within the educational sector. According to Bryson (2018), these models increasingly integrate decision maker's behavioral dynamics, drawing upon frameworks such as Ajzen's theory of planned behavior to better navigate institutional complexities and environmental variables. This evolution marks a transition from traditional to contemporary planning approaches, emphasizing the critical role of strategic, cultural, and political dimensions in effective resource allocation and strategic alignment, especially in higher education.

### *Advancement of Planning Models*

Planning models have evolved over time from traditional budgeting models to more integrated and strategic planning models (Koteen, 1997). Traditional budgeting models were often incremental and focused on short-term goals, whereas more recent planning models have emphasized long-term strategic planning and resource allocation (Bryson, 2017). They highlight the importance of inclusiveness and openness in planning processes and the need for meaningful planning outcomes that align with an institution's mission and objectives (Barr & McClellan, 2018). De Lancer Julnes et al. (2020) highlighted the importance of cultural and political factors (e.g., leadership, collaboration) in driving and sustaining performance measurement and community indicators integration efforts. The advanced planning models suggest paying close attention to cultural and political factors, such as leadership and collaboration. Recognizing the evolved nature of planning models, it is important to learn more about specific strategic planning models.

### Strategic Planning Models

Effective strategic planning is vital for any educational institution, and various models have been proposed in literature to guide this critical process (George, 2017). One of the most used methods is SWOT analysis, which stands for strengths, weaknesses, opportunities, and threats. This model aids in institutions in identifying internal and external factors that can influence their strategy, enabling them to build on strengths, remedy weaknesses, capitalize on opportunities, and mitigate threats (Allison & Kaye, 2005; Bryson, 2004).

Another model that has gained prominence in literature is the rational planning model, detailed by Bryson (2004). This model emphasizes a logical approach to planning, beginning with a clear understanding of organizational goals, followed by a comprehensive analysis of the various courses of appropriate action to achieve these goals. The model provides a structured framework that assists organizations in making decisions based on a thorough understanding of their status and future objectives (Hovey, 2012).

Whereas these models provide an excellent starting point, the evolving environment in which community colleges operate necessitates the incorporation of additional variables. Economics and external data are increasingly important for community colleges to consider in their planning efforts. Factors such as local unemployment rates, GDP growth, and changes in fiscal policy can have a direct impact on funding levels and, consequently, the resources available for achieving strategic goals (Goldberg & Prottas, 2017; Hovey, 2012).

Integrated planning and budgeting offer a harmonized approach to strategic planning and financial management (J. Williams, 2000). Goho and Webb (2003) explained administrators should consider the annual budget as the "financial plan to accomplish the strategic plan" (p. 384). In this view, strategic planning guides operational planning, and both are intrinsically linked. This integrated approach ensures the institution's resource allocation aligns with its broader goals and objectives, allowing for more effective and efficient operations (Salem et al., 2020; J. Williams, 2000).

**Leaders' Ability to Allocate Resources Strategically**

The success of any educational institution depends on the efficacy of its leadership in making informed and rational resource allocation decisions (Boggs & McPhail, 2016). The practice of integrated planning serves as the best approach in this regard, matching strategy with budget allocation to navigate organizational goals (Salem et al., 2020; J. Williams, 2000). Integrated planning ensures budgeting is not an isolated process but tied to the institution's strategic requirements (J. Williams, 2000). This practice aligns the allocation of resources like personnel, capital, and technology with an organization's strategic plan, ensuring priorities are financially supported (Allison & Kaye, 2005; Lorenzo, 1993).

However, implementing integrated planning has its challenges. Lorenzo (1993) highlighted leaders' struggles with the complexity and length of the strategic planning process. Leaders often need help with devoting time to careful planning and the need for agile action. They value strategic planning but express impatience with its time-consuming nature (Bryson, 2017). This importance highlights the significance of planning models that not only offer strategic insights but are also operationally practical. Leaders want models that help them act quickly while maintaining flexibility to adapt to changes, which is vital given the financial uncertainties that community colleges often face (Lorenzo, 1993).

Allison and Kaye (2005) also touched upon the importance of actionable planning in their work on strategic planning for nonprofit organizations. Although their focus is not specifically on educational institutions, the parallels are hard to ignore. They advocate for strategic plans that are realistic, achievable, and aligned with resource allocation. This approach reduces the gap between strategy formulation and implementation, making it easier for leaders to allocate resources consistent with organizational goals (Allison & Kaye, 2005).

Adding to the complexity of strategic resource allocation is the inherent politics involved in decision-making processes (Mappadang et al., 2021). Bryson (2004) posited the political process yields consensus on only a few topics that resonate as overriding concerns for various stakeholder groups. In community colleges, this could mean leaders must navigate conflicting interests, such as faculty demands, administrative needs, and state or local government directives when allocating resources (D. Williams, 2004). This political balancing act further emphasizes the need for strategic planning models that offer leaders a robust framework for making difficult choices. When a leader can effectively negotiate these political landscapes, it enables the college to focus its limited resources on strategically important areas, thereby offering the best chance of achieving institutional goals (Bryson, 2004). With the discussion over the complexities of leadership and resource allocation, it becomes increasingly clear that effective planning is only one side of the coin. The other essential aspect is budgeting, which is the financial design for executing these plans (Barr & McClellan, 2017).

**Budgeting**

Effective planning and budgeting are inextricably linked, serving as the backbone for any educational institution aiming for fiscal responsibility and strategic alignment (Allison & Kaye, 2005; Bryson, 2004; Goho & Webb, 2003). Whereas planning outlines the "what" and "why," budgeting delineates the "how," specifying the resources required to achieve set objectives (Elbanna et al., 2016). However, for budgeting to effectively support the institution's mission and vision, a fundamental review of the budget process and allocation decisions is essential (Israel & Kihl, 2005). In this context, it becomes necessary to ensure that base allocations are flexible and not merely historical events that may resist change. Adhering to fixed allocations with regular review may help the institution adjust to new priorities or challenges, therefore, failing to reflect the current mission and vision (Salem et al., 2020).

Several adapted processes have been suggested in the literature to enhance the efficacy of budgeting. One such approach is zero-based budgeting, where every expenditure must be justified for each new period, eliminating the baseline of ongoing costs from past budgets (Allison & Kaye, 2005). Another is the contribution margin approach, which focuses on generating and analyzing per-unit metrics to guide allocation decisions (Bryson, 2004). These methods promote a culture of continuous improvement and accountability, aligning resources more closely with strategic goals (Rabovsky, 2012).

Furthermore, a budgeting process characterized by inclusiveness and openness invites broader participation from various stakeholders, including faculty, staff, and students (Goho & Webb, 2003). An inclusive process not only democratizes decision making but also adds layers of perspective that might otherwise be missed. It is a step toward creating a shared understanding and consensus around budgetary decisions, fulfilling the political function of bringing varied interests to a common ground (Bryson, 2004).

State funding helps make higher education accessible to students who may not be able to afford it otherwise, and it is critical for maintaining the quality of higher education. State-funded higher education institutions play a vital role in economic development (Rabovsky, 2012). Higher education institutions provide a skilled workforce for businesses, attract research funding, and create innovation and entrepreneurship opportunities (Pasadena City College, 2002), which is the return on states' investment in higher education.

Husted and Kenney (2019) indicated state appropriations had no statistically significant impact on budgeted research spending. However, Pratolo et al. (2020) stated management competence and reward systems positively influence performance-based budgeting implementation. Moreover, performance-based budgeting implementation plays a significant role in higher education institutions' quality (Orr & Usher, 2018). Specifically, implementing performance-based budgeting can enhance organizational performance. Using a budget system based on institutionalized, historical levels has been common in many organizations. Resource allocation decisions must be linked to an organization's mission and planning priorities to achieve organizational goals and objectives (Bryson, 2017), meaning budget decisions must be made based on the strategic direction of the organization and the priorities identified in its planning process (Salem et al., 2020). Misallocation of resources can hurt the organization's ability to achieve its goals. Therefore, organizations must go under transition from the old budget system. Instead, they must adopt a budget system aligned with their mission and planning priorities. This adoption allows them to construct informed decisions about the allocation of resources (Bryson, 2017). Aligning budget systems with organizational missions and goals requires focus shifts to integrated planning and budgeting. This approach emphasizes the importance of strategic planning in guiding not only the allocation, but also the reallocation of resources in an institution, ensuring financial decisions are forward looking and aligned with broader organizational strategies and environmental considerations.

**Integrated Planning and Budgeting**

Strategic planning should guide a college's administrative activities in an integrated planning and budgeting process, which also leads to guide resource allocation and reallocation (G. Cohen, 2009). Jimenez (2014) revealed rational analytic techniques, such as strategic planning and performance management, can help local governments make targeted cuts in expenditures during periods of fiscal decline. Doing so can make them leaner and more effective while preserving administrative capacity (Hovey, 2012). Moreover, budget predictions provide a basis for planning and decision making (Popesko et al., 2016). Organizations can develop realistic budgets and allocate resources effectively by forecasting future revenues and expenses (Barr & McClellan, 2017). Budget predictions are critical for maintaining financial stability (Rabovsky, 2012). When organizations do not predict their future revenues and expenses, they may find themselves facing unexpected financial challenges (e.g., cash flow problems, debt issues, the need to make sudden cuts in services). Popesko et al. (2016) showed a negative relationship between GDP fluctuations and a firm's ability to predict changes in a business environment. Specifically, higher GDP fluctuations lead to a lower ability to predict changes in a business environment (Popesko et al., 2016). Such a relationship highlights the significance of integrating strategic planning and budgeting.

Bergmann et al. (2020) found integrating analytical methods into the budgeting process can lead to increased satisfaction with it. Scholars often criticize traditional budgeting systems for needing to be more time consuming, costly, and inflexible (Ekholm & Wallin, 2000). However, modern analytics may overcome these problems and lead to increased satisfaction with the budgeting process. From an organizational management perspective, business analytics impacts managerial accounting from an enterprise systems and business intelligence perspective (Appelbaum et al., 2017). Such research has also indicated the role of management accountants has significantly changed over the years. Modern management accountants work from four aspects: (a) to participate in strategic cost management to achieve long-term goals, (b) to implement management and operational control for corporate performance measures, (c) to plan for internal cost activity, and (d) to prepare financial statements (C. Drury & Triest, 2013). Numerous colleges have employed an incremental budget system, which typically solidifies funding levels for program and service areas at previous levels and only makes new funding decisions when additional resources arrive (CCCCO, 2022; Frost, 2009). The need to connect resource allocation choices to mission and planning priorities necessitates discontinuing a budget system based on institutionalized, historical levels. Given the integral role that budgeting plays in aligning resources with strategic goals, the next logical step is to discuss the planning process itself. This process is pivotal for community colleges, as it leads to meaningful outcomes that should be aligned with the institutions' missions.

**Planning Process**

In the complex landscape of community colleges, having a meaningful planning outcome is required to align with institutional objectives. One of the pillars of strategic planning in educational settings is to ensure alignment with the mission of community colleges, which often encompasses goals like providing accessible education and contributing to workforce development (Allison & Kaye, 2005; Bryson, 2004).

The concept of meaningful planning outcomes is closely tied to the institution's mission (Allison & Kaye, 2005). For example, when budgeting and resource allocation align with the mission, the planning outcome transcends mere financial management and becomes a strategic action for fulfilling the institution's core objectives (Israel & Kihl, 2005). This form of alignment provides a framework in which resources can be effectively distributed to maximize educational impact and community reach (Goho & Webb, 2003).

To achieve a meaningful planning outcome, the planning process must be intrinsically mission driven (Lorenzo, 1993). McAndrew (2023) found mission-driven organizations perform more effective planning approaches than non-mission-driven organizations. Moreover, organizations that employ detailed planning reported better strategic alignment and faster growth, indicating integrated strategic planning mechanisms significantly contribute to organizational performance (Kaplan et al., 2001). Therefore, every aspect of strategic planning, from budgeting to resource allocation to program development, should aim to further the mission of the community college (Bryson, 2004). Such alignment not only facilitates decision making but also promotes a greater sense of purpose among stakeholders, emphasizing the key role of community colleges in the broader educational landscape (Phelan, 2005).

**Forces of Change: Internal and External Factors**

Given the fundamental role of mission-driven planning processes in community colleges, it is essential to consider the various changes affecting both planning and outcomes. This section discusses these influential factors, which include social and demographic shifts, technological advancements, revenue volatility, and increased accountability. Leaders responsible for implementing an integrated planning process that connects resource allocation decisions to the accomplishment of mission-oriented initiatives are working in a situation where previous assumptions about operating costs, instructional delivery methods and technologies, enrollment growth, and the makeup of the student body may no longer be accurate. The students and communities who colleges serve are being influenced by social and demographic factors (California Community College Chancellor's Office [CCCCO], 2024; A. Cohen & Brawer, 2003). Continual advancements in technology need the provision of current training in new job-related abilities and the need to revamp service provision and administrative procedures (Boggs & McPhail, 2016; CCCCO, 2024; A. Cohen & Brawer, 2003). These forces of change necessitate a substantial allocation of resources and result in ongoing expenses across all operational domains, whereas revenue streams continue to be unpredictable (A. Cohen & Brawer, 2003; Dougherty & Natow, 2015).

### *Social and Demographic Changes*

Social and demographic changes significantly influence the planning process in community colleges (Allison & Kaye, 2005). For example, a rise in enrollment from diverse populations may require reallocating resources to specialized programs or support services (Phelan, 2005). These changes require flexible planning strategies to adapt to a dynamically evolving student body and community needs (Lorenzo, 1993). Community colleges have experienced an expansion in their mission. At the same time, significant social, economic, and demographic changes have occurred, resulting in a transformation of their communities and, consequently, their student populations. Societal factors (e.g., the dissolution of the family as a social institution, the rising rates of poverty among young people) are exerting greater demands on institutions to offer tailored programs and services (Boyd, 2002; Cohen & Brawer, 2003; Kozeracki & Brooks, 2006). Furthermore, these issues contribute to the rising rates of part-time attendance and the growing population of entering students who require more essential skills (A. Cohen & Brawer, 2003; Kozeracki & Brooks, 2006).

The demand for higher education is increasing due to the expanding needs of younger, traditional students and the growing number of "baby boomer" children and first-generation students. This trend is expected to continue in the next decade, resulting in a more significant percentage of the U.S. population pursuing education beyond high school (Boggs, 2003; Bryson, 2004). Many immigrant students are characterized as first-generation college students and nonnative English speakers (CCCCO, 2023; Scott & Stanfors, 2010) and come from a wide range of culturally varied backgrounds. The ongoing economic uncertainty, along with significant changes in the workplace that impact the skills and competencies desired by employers, has resulted in a simultaneous increase in the number of adults who are returning to community colleges for retraining (Callan, 2009; Hendrick et al., 2011). These immigrant students and older adults constitute a burgeoning group with distinct educational and support requirements (Bryson, 2004; A. Cohen & Brawer, 2003). The shifts in the student population necessitate the development of more all-encompassing course options and enhanced student support services (CCCCO, 2023; Kozeracki & Brooks, 2006). The lack of diversity among college instructors and staff, who often need to understand the needs, cultural background, and diversity of their students, exacerbates the difficulty (A. Cohen & Brawer, 2003). To ensure that all students can effectively participate in a thriving learning environment, it is imperative to strengthen and enhance support services and provide primary training for teachers and staff (Beyer & Ruhl-Smith, 1999).

### Technological Changes

Technological advancements also play a pivotal role. The ongoing advancement of technology and its use in education and student support adds complexity to efficiently allocating resources to fulfill the college's objectives. Colleges must allocate annual budget funds to ensure that classrooms have modern equipment and to keep up with the increasing demand for innovative distance learning methods (A. Cohen & Brawer, 2003). Simultaneously, online education and traditional instruction must be clarified as instructional technologies are integrated into all courses (CCCCO, 2010; A. Cohen & Brawer, 2003). Student support services are now available in digital format to better assist the diverse student population and cater to their individual needs (Accrediting Commission for Community and Junior Colleges, 2014; CCCCO, 2010). With the increasing implementation of technology across the college campus, there is a corresponding increase in spending on equipment and infrastructure to ensure that instruction and services remain current. The continuous updating of applications and implementation of new technologies necessitates a significant investment in comprehensive professional development for academics and staff, which incurs an additional expense that needs to be paid.

Colleges face the difficulty of offering current technical training in career technical programs, ensuring education remains applicable as job roles evolve, and incorporating increasing amounts of technology in the industry (Boggs & McPhail, 2016; A. Cohen & Brawer, 2003; Nunley et al., 2011). According to others, the expensive upkeep of career technical programs has hindered community colleges from fulfilling their other goals, such as offering possibilities for students to transfer to 4-year institutions (Dougherty, 2001; Jacobs & Dougherty, 2006). Given that career technical education programs are often more expensive and have lower enrollment than college transfer options, college presidents must strategically manage funds to ensure the success of their institutions in the face of declining funding.

However, the high prices of advanced technology are not exclusively caused by career technical programs. According to the CCCCO (2010) Data Mart, 4-year colleges and universities now require college transfer students to have a greater understanding and proficiency in technological applications. This includes general education courses and their selected field of study. Proficiency in specialist applications and equipment in subjects such as biology, statistics, and library science are a requirement in all advanced schools and, increasingly, for all beginner-level professions. In addition, many older entering and basic abilities students need to gain the requisite technical abilities to excel in college-level courses. These students require instructional support in fundamental computer operations, as The Research & Planning Group for California Community Colleges (The RP Group) highlighted in 2007.

The challenge that technological development presents to planners and college administrators encompasses a broader range of services and instructional delivery methods, bringing about additional expenses and necessitating further training for teachers, staff, and students. The compounding impact of these environmental changes is to amplify the already daunting task that college leaders will encounter in the future when it comes to allocating college resources to address various demands. The dynamic nature of the current environment puts pressure on schools to adapt fast and stay focused on their mission priorities. This pressure has led to a shift toward implementing efficient and adaptable planning processes (Allison & Kaye, 2005; Boggs, 2003, 2007; Bryson, 2004; Goho & Webb, 2003).

### Revenue Volatility

Revenue volatility is another significant factor affecting planning in community colleges (Israel & Kihl, 2005). Many similar changes that have influenced community colleges have also influenced public services. The demand for education and other public services varies by economic cycles and demographic shifts. An increasingly aging population necessitates more investments in healthcare services while rising enrollments necessitate additional funding from state educational resources. Community colleges across the country increasingly depend on state money as their main source of revenue (Boyd, 2002; Hovey, 1999). As a result, they are now in competition with other public education providers and service sectors for a greater portion of the same state funds. The funding history of community colleges partially roots the tension around state funding for public education entities. Initially, community colleges relied heavily on tuition fees and local community revenue sources for support (Pedersen, 2005). Before the 1960s, state legislatures opposed any efforts to obtain state money because they wanted to restrict the expansion of local schools and ensure that local communities could provide for secondary schools (Pedersen, 2005). State universities also expressed opposition due to concerns about decreasing enrollments in their schools. They were worried the decrease in enrollments would lead to a drop in revenue, which is necessary to support ongoing fixed and operating costs (Pedersen, 2005).

The formal subsidization of community college operations by states occurred in the 1960s when federal student aid programs began offering offsets for community college tuition. Although the enrollment trend has been downward recently, community colleges experienced substantial growth by enrolling subsidized students, thanks to the federal student loan programs that served as a large source of additional funding. Consequently, community colleges started vying with state-funded universities and public K–12 schools for the same funding source (Pedersen, 2005).

In the service industry, competition for state budget funds is also increasing. The proportion of the nation's population aged 65 and above is increasing quicker than other age groups. This demographic trend is attributed to the aging of the baby boomer generation, the largest generation in U.S. history (Boyd, 2002). The demographic shift toward an older population indirectly affects the allocation of funds for community colleges. The increase in Medicaid expenses, especially for nursing home care, necessitates a larger portion of the state budget (Boyd, 2002; Pedersen, 2005). With an increasing number of baby boomers entering retirement, Medicaid expenses will continue to rise. This growing demand will increase state expenditure on healthcare at a higher rate than the growth in state revenue (Boyd, 2002; Hovey, 1999). The demand for higher education services is outpacing the expansion in state and municipal tax revenues. Hovey (1999) asserted a persistent structural state budget deficit reduces state funding for community colleges and other higher education branches over the next 10 years. Hovey's discoveries formed the foundation for the following report by Boyd, who employed comparable techniques to create an 8-year forecast of state revenues nationwide.

The economic recession of 2008, characterized by a significant decline in property values and elevated unemployment rates, has profoundly affected state budgets. Decreased property tax values and personal income levels reduced state governments' revenue sources, whereas there was an increase in spending on public services and unemployment compensation. The current structural deficits are consequently more substantial, as they should have considered the possibility of another economic recession in their estimates. However, shifting demographics and societal dynamics persist, and people may anticipate ongoing uncertainties in funding levels for the foreseeable future. Based solely on demographic factors, Hovey (1999) and Boyd (2002) argued higher education sectors will require more funding to sustain the current level of services for the growing student population. This finding does not even consider the evident need for diversification in services and educational offerings. As anticipated by all predictions, the strain on state funding sources will result in persistent structural budget deficits and lead to a more unpredictable and unstable flow of finances. Amidst the pressure to fulfill all aspects of the community college mission, the ongoing and repetitive funding cuts and the possibility of sustained allocation reductions for a decade or more necessitate cost-saving measures and reductions in the level of services and quantity of instructional offerings (Tirol-Carmody et al., 2020).

A survey of 120 community college presidents and chancellors revealed how these institutions handle the ongoing economic recession and limited financial resources at the federal and state levels (Green, 2009). Several presidents expressed they were confronting the same issues they had faced 6 years prior, during the period when community schools struggled to recover from the 1990s economic decline and faced the recession in 2003 (Olaode, 2015). According to Green (2009), the presidents in 2009 employed familiar short-term tactics to decrease spending and manage reductions in financing, similar to what they had done in prior economic downturns. These measures encompassed enacting a moratorium on new hires, cutbacks in travel and discretionary spending, and vigilant oversight of all financial outlays. During a typical economic recession, colleges cannot maintain these interim expenditure reductions for a prolonged period without significantly impairing their operational efficiency. Colleges facing budget cuts in recent years have seen an average reduction of approximately 5% in financing. The additional impact of midyear budget returns has further exacerbated this issue. Despite the financial limitations, 92% of college presidents stated they observed a rise in the number of students enrolled, with 70% reporting an increase of over 5% (Green, 2009). Consequently, although presidents responded to budget reductions by implementing freezes and restrictions, they faced challenges in advancing program development and expansion to accommodate the growing student population (Green, 2009).

***Accountability***

Simultaneously, college administrators have a duty to align funding with mission objectives while also experiencing pressure from accrediting organizations and government funding sources to provide evidence of student success and progress (CCCCO, 2024). Concerns about education quality and the United States' global competitiveness have led to demands for greater accountability and the adoption of efficient assessment systems across all areas of education. Various variables and the increasing complexity of their purpose now obligate community colleges to provide evidence and report on student progress and performance. The evaluation of student progress and performance includes demonstrating the gradual attainment of student learning outcomes in their program (CCCCO, 2024). According to reports from the federal Department of Education and the American Association for Community Colleges, community colleges in the United States are generally successful in providing access to education. However, a significant proportion of students fail to achieve their educational goals, as indicated by data from the National Center for Education Statistics in 2010. The success and persistence rates of CCCs are like those of the entire nation, as reported in (CCCCO, 2024). According to a report by the state Chancellor's Office, just 52% of California community college students who started their education were able to achieve their goals after 6 years of enrolling (CCCCO, 2010, 2024). Student success measures have suggested there is considerable scope for enhancing student performance in community college remedial courses. According to the California Community College Basic Skills Accountability Report (CCCCO, 2010), the majority of entering students who were assessed as needing remedial work in mathematics between 2005 and 2008 tested at three or more levels below the level of preparation required for college. Out of this greater percentage of students enrolled in basic skills math, only 12% eventually achieved success in finishing a mathematics course at the transfer level. The report also revealed most students assessed as having English abilities below the college level performed at a level more than two levels below the college level during the same timeframe (CCCCO, 2010). According to the CCCCO (2010), around 35% of students who need more than two college preparatory English sessions were able to successfully finish a transfer-level English course after completing the basic skills sequence.

The current focus on accountability has created demands for community colleges to enhance their performance in serving students. Nevertheless, these accountability demands have also imposed further financial burdens on the colleges. It is necessary to create assessment procedures and establish data collection systems. It is necessary to provide technical applications and infrastructure to facilitate data evaluation and examination, as well as to recruit and train personnel to manage the systems. Procedures for accountability aim to bring about educational reform by considering the distribution of resources and investing in new training, programs, and services. The increasing emphasis on accountability adds pressure to community college finances and necessitates thorough evaluation and strategic planning.

**Leadership**

These changes—whether they are social, technological, financial, or regulatory—shape the context in which community colleges operate. Understanding the array of changes that impact community colleges sets the stage for discussing the vital role of leadership in navigating these complexities (Boggs & McPhail, 2016; Kotter, 1996). Administrative and financial capacity, leadership support, and goal clarity are positively associated with the use of strategic management (Dimitrijevska-Markoski et al., 2021). Moreover, Alharafsheh et al. (2021) examined the mediating role of strategic planning by analyzing the relationship between entrepreneurs' characteristics and strategic performance, with and without the presence of strategic planning as an intermediate variable. The results showed strategic planning partially mediated the relation between entrepreneurs' characteristics and strategic performance, indicating it plays an essential role in improving the performance of Jordanian private universities. The next topic emerges from the link between leadership and decision making. Environmental scanning is essential for any strategic planning effort that involves leaders systematically exploring internal and external environments to gain insights vital for informed decision making.

**Environmental Scanning**

Understanding the internal and external environment is a key part of strategic planning for any organization, including community colleges (Park et al., 2017). Environmental scanning is a critical preliminary step to help institutions develop a comprehensive understanding that informs subsequent goal setting and planning activities (Allison & Kaye, 2005; Bryson, 2004). Over the years, the scope of environmental scanning has evolved considerably, moving beyond an internal focus to include many external factors that significantly influence organizations (Lorenzo, 1993).

Initially, environmental scanning in business and industry sectors centered primarily on internal aspects like human capital, resource costs, and organizational capacity (Lorenzo, 1993). However, the rise of globalization and the increased interconnectedness of economies have led to a shift (Bolman & Deal, 2021; Lee & Terrence, 2021). Today, external elements such as global trends, economic fluctuations, and social changes are equally, if not more, important in shaping strategic decisions (Lorenzo, 1993). It is particularly relevant for community colleges in California, where budgeting techniques and funding levels often need to align with institutions' strategic goals, leading to ineffective resource allocation (Hovey, 2012).

In community colleges, economic indicators like unemployment rates and GDP growth influence funding levels, making them critical variables in the environmental scanning process (Hovey, 2012). Similarly, understanding fiscal policies such as those affecting income, corporations, sales, and local property taxes is integral to predicting and preparing for changes in funding (Bers & Head, 2014; Garrett et al., 2023; Hovey, 2012). Effective environmental scanning thus requires a multifaceted approach that blends both economic and policy aspects, offering a composite view that helps in more accurate forecasting and planning (Phelan, 2005).

Economic factors do not merely influence community colleges; they are also significantly affected by social and demographic changes. Population growth, or the lack thereof, can heavily impact enrollment numbers, thereby affecting budgeting and resource allocation (Barr & McClellan, 2017).

Additionally, technological advancements offer both opportunities and challenges; for example, the move toward more online education can impact costs positively and negatively (Bryson, 2004).

The increasing focus on accountability also presents another layer to the environmental scanning process. The demand for greater transparency in the allocation and use of resources has necessitated incorporating accountability measures into strategic planning (Dimitrijevska-Markoski et al., 2021). Hence, an effective environmental scan not only looks at economic indicators and fiscal policies but also considers accountability standards that align with the institution's strategic objectives (Hovey, 2012; Lesca, 2011). Environmental scanning has become an essential tool in the modern planning process for community colleges (Hovey, 2012). A well-executed scan incorporates a diverse range of factors, from economic indicators to fiscal policies to social and technological changes, helping these educational institutions to adapt and thrive in an increasingly complex environment (Lesca, 2011). Mustika and Rahmayanti (2019) found that various factors, including political, legal, economic, social, technological, and environmental aspects, significantly influence the strategic planning and implementation of inclusive education for individuals with different abilities. Moreover, engaging in environmental scanning can benefit the planning process for academic programs by helping to identify threats and opportunities in the professional, competitive, institutional, and higher education environments, leading to more informed decision making and strategic goal setting (Strubhar, 2011). Contingency forecasting serves as a critical extension, explicitly focusing on the impact of economic and demographic shifts on funding levels (Chermack, 2011). This specialized form of planning fills a gap that traditional environmental scans may need to pay more attention to, particularly the challenges posed by revenue volatility. As environmental scanning provides a broad understanding of various factors affecting community colleges, the focus inherently shifts to contingency forecasting. This specialized approach in planning takes a deeper dive into predicting and preparing for potential economic and demographic shifts, particularly emphasizing the critical aspect of funding stability and the impact of revenue volatility.

**Contingency Forecasting**

Contingency forecasting refers to a planning process that focuses on understanding how economic and demographic changes may affect an institution's ability to maintain funding levels (Barr & McClellan, 2018; Chermack, 2011). Traditional environmental scanning focuses on factors that change enrollment and industry demand for services and programs (Lesca, 2011). The traditional environmental scan is a planning process that identifies internal and external conditions expected over a planning period. We must understand how economic and demographic changes may affect the state government's ability to maintain funding levels. Many scans ignore revenue volatility's impact on an institution's ability to plan for the future (McKeown-Moak & Mullin, 2014). Contingency forecasting is essential for community colleges to adapt and thrive in a changing environment. The consensus among top executives is that effective resource procurement and management are essential for any organization's success, highlighting the importance of strategic planning in leadership (Bragg, 2011). Contingency forecasting enables college leaders to forecast revenue changes, providing them with the desired skill and resilience while ensuring the planning processes remain consistent, which is particularly critical during periods of economic hardship (Barr & McClellan, 2018).

**Conceptual Framework**

A conceptual framework should be a component that guides the research. It provides a structured approach to understanding, exploring, and analyzing the research problem (Creswell, 2014). In essence, a conceptual framework serves as a map or blueprint for the dissertation, guiding every aspect of the research process, from developing research questions to interpreting results. It helps ensure coherence and rigor in research, making the study more robust and credible (Creswell, 2014; Johnson & Onwuegbuzie, 2004). This study focused on the influences of economic factors on changes in the annual revenue of community colleges, which then influences and gets shaped through the colleges' strategic plan.

Pragmatism and RDT are the philosophical foundations and the theoretical framework that define why external conditions impact integrated strategic planning. RDT argues organizations rely on external resources to achieve objectives and commit to their missions. To use these resources, they must interact with and adapt to their external environments (Hillman et al., 2009). Consequently, these external entities and conditions can influence an organization's behavior and decision-making processes. Pragmatism is a critical foundation of integrated strategic planning where leaders test ideas to promote effectiveness and efficiency, and pragmatists are more interested in finding practical solutions than developing abstract theories (Dennes, 1940).

Pragmatism functioned as the overarching philosophical framework that directs the focus of this study toward the pragmatic implementation of budgeting methodologies and strategic planning in CCCs, with the aim of aligning them with their mission (see Figure 1). The conceptual framework outlined began with economic indicators, suggesting their pivotal role in shaping the budgetary considerations of community colleges. Subsequently, there was a gradual development toward the creation of a comprehensive and efficient strategic plan, a procedure deeply rooted in practical philosophy. Pragmatism promotes a method of problem solving that involves taking action and making adjustments based on the results obtained in real-world situations. This ideology served as the foundation for the planning process, guaranteeing the colleges' strategic responses to external economic fluctuations are both feasible and adaptable. The last component of the framework demonstrated how these strategically informed plans directly help to achieving the main objective of community colleges, defined by the CCCCO. This in-depth examination highlighted the importance of pragmatics in connecting economic realities with strategic objectives, allowing educational institutions to successfully manage financial issues while actively pursuing their educational aims.

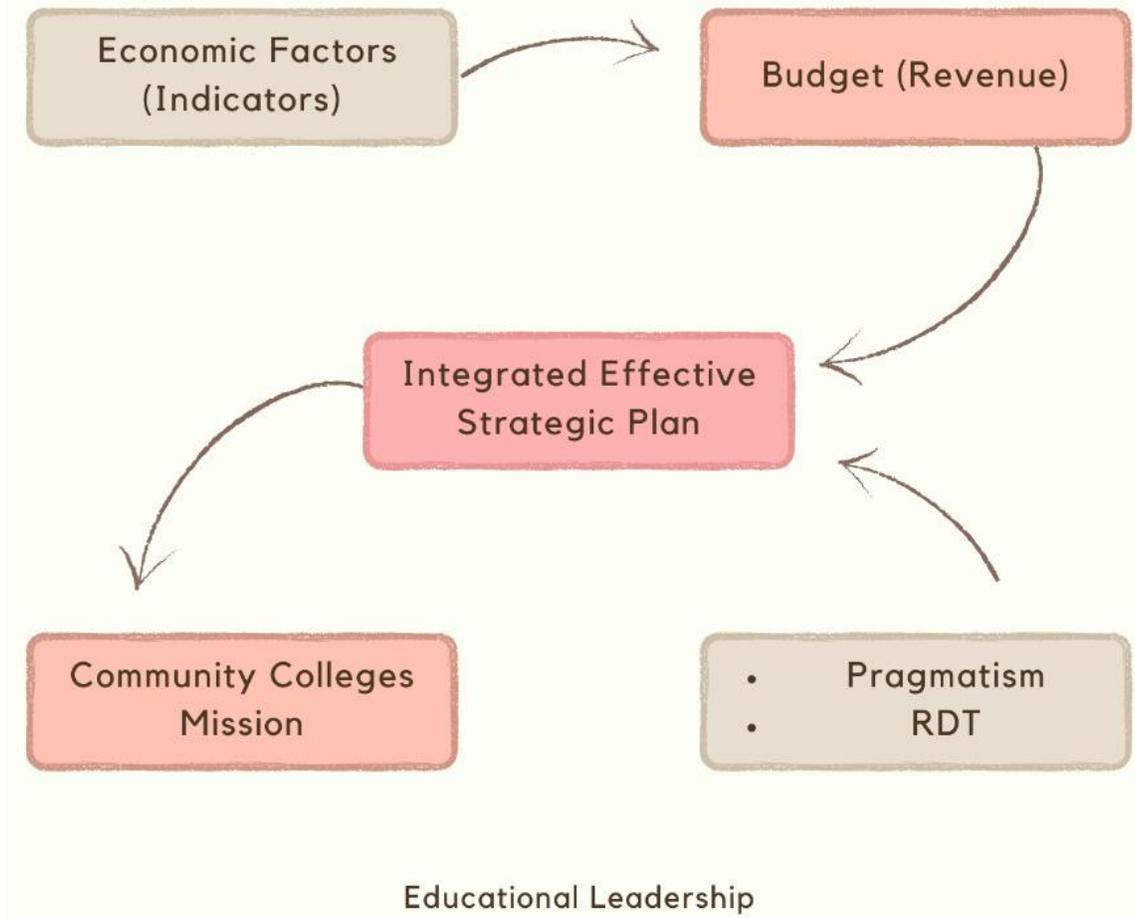

*Figure 1.* Conceptual framework, the connection between environment and community colleges mission.

**Chapter Summary**

In conclusion, the existing literature has extensively discussed various aspects of strategic planning, budgeting, and resource allocation in educational settings, including community colleges. Researchers have proposed different planning models and emphasized the importance of aligning these models with effective budgeting techniques to achieve institutional goals (Allison & Kaye, 2005; Bryson, 2004; Goho & Webb, 2003). Whereas existing research has addressed various aspects of strategic planning models, leaders' abilities to allocate resources, and different budgeting techniques, there remains a significant gap in understanding how these elements interact with economic indicators like unemployment rates and GDP growth, especially in the context of financial management in CCCs. This study aimed to fill this gap by investigating the correlation between economic indicators and funding levels, as well as the predictive value of these variables for future budget decisions in the specific context of CCCs. Building upon the comprehensive review of strategic planning, budgeting, and resource allocation discussed in Chapter 2, Chapter 3 shifts to the methodological approach of this study. It details the research design, data collection, and analytical techniques employed to explore the interaction between economic indicators and funding levels in CCCs, directly addressing the identified gaps in existing literature.

# CHAPTER 3

# METHOD OF INQUIRY

Efficient resource allocation and financial accountability are vital in higher education, especially in community colleges (Salem et al., 2020). The Accrediting Commission for Community and Junior Colleges (2024) has established rigid requirements for these areas to assure a standard of academic integrity. Nevertheless, many colleges struggle with effective resource management, which leads to concerns about compliance with regulatory standards and operational effectiveness, especially in budget management (Barr & McClellan, 2017). Such an issue impacts higher education institutions mission and responsibilities. The problem this study addressed was the concern that arises when the use of advanced budgeting techniques in community colleges do not align with the institutions' mission and efficient budget allocation. To address this problem, the study investigated relationships between economic variables and funding levels and the predictive value of these variables for future budget decisions. This study answered three questions:

1. What is the relationship between the advanced budgeting techniques (economic indicators such as unemployment rates and GDP growth) and funding levels for California Community Colleges?

2. What is the relationship between fiscal policies (those impact on income, corporation, sales, and local property taxes) and funding levels for California Community Colleges?

3. How do the relationships identified in Research Questions 1 and 2 predict future budget allocations?

This chapter outlines the quantitative methods I used to examine financial management practice at California Community Colleges (CCCs). This chapter details the study's sample, survey instrument, data collection procedures, variables, and statistical tests.

## Quantitative Research

Scholars define quantitative research methodology as a systematic empirical investigation of observable phenomena via statistical, mathematical, or computational techniques (Creswell, 2014). Quantitative research focuses on creating and using mathematical models, ideas, and phenomena-related hypotheses (Creswell, 2014). In quantitative research, measurement is critical because it establishes the essential link between empirical observation and the mathematical representation of quantitative connections (Shields & Rangarajan, 2013). The philosophical foundation of quantitative research lies primarily in positivism, which asserts that knowledge is based on natural phenomena and their properties and relations as verified by empirical sciences. Positivist researchers believe they can observe and describe reality objectively without interfering with the studied phenomena. They focus on facts, and their research is often top down, guided by theory or hypotheses. Quantitative approaches enable numerical data collection and objective analysis (Creswell, 2014), and quantitative tools align with the numerical nature of financial data and allow me to draw data-driven conclusions (Leedy et al., 2019).

Quantitative methodology, although robust in many aspects, does have specific weaknesses. One fundamental limitation is its potential lack of depth regarding complex human behaviors and attitudes (Creswell & Creswell, 2018). Quantitative methods often prioritize breadth over depth, leading to a scenario where the emphasis on statistical analysis and generalization can overshadow the richness of data (Leavy, 2023). Another limitation is its reliance on preestablished hypotheses and models, sometimes overlooking novel or unexpected findings that don't fit the theoretical framework. Moreover, quantitative research can sometimes be too rigid due to its structured nature, potentially missing nuances that more flexible, qualitative methods might capture (Creswell & Creswell, 2018; Leavy, 2023).

Concerning this study, these weaknesses manifested as a need to fully capture the nuanced relationships between budgeting practices and educational outcomes or as missing out on unanticipated factors influencing these relationships. At the same time, such data may need more clarification on the underlying reasons behind why specific misalignments in budgeting and strategic objectives occur. Researchers select quantitative research methods because they can produce objective, generalizable, and testable data (Creswell, 2014), and this method was the best design for this study as it provided the most valuable data on correlations and trends.

**Research Design**

To effectively resolve the research issue, researchers in quantitative research select a research design that integrates different study components in a cohesive and logical manner. Research design forms a blueprint for data collection, measurement, and analysis (Creswell, 2014). Common designs in quantitative research include experimental, correlational, and survey designs. According to scholars, experimental designs have control over the variables and use random assignments, whereas correlational designs do not involve manipulation or control of variables (Creswell & Creswell, 2018). Scholars have categorized experimental designs as having control over the variables and using random assignment, whereas correlational designs do not involve manipulation or control of variables (Creswell & Creswell, 2018).

Correlational design is a nonexperimental research method frequently employed in quantitative research to evaluate any relationship between two or more variables without modifying them (Creswell, 2014; Leavy, 2023). This approach seeks to determine the direction and magnitude of the link between variables. Correlational research entails observing variables in their natural environments and quantitatively analyzing their links or associations. The correlational design does not prove cause-and-effect linkages; it only suggests the presence of a relationship (Trochim & Donnelly, 2008).

Selecting a correlational research design for this study was highly justified given the nature of the research questions, which focuses on exploring the relationships between economic data over time and fluctuations in the state budget for colleges. Creswell and Creswell (2018) and Trochim and Donnelly (2008) stated correlational research design is particularly well suited for investigating these relationships because it allows for analyzing how variables change together without manipulating the variables themselves. Because this study aimed to examine existing data sets to identify patterns and associations, a correlational design was the most appropriate choice. It enabled me to assess the strength and direction of the relationships between economic indicators such as unemployment rates and gross domestic product (GDP) growth and changes in funding levels for CCCs.

Furthermore, a correlational design aligned well with the scope of this study, which was to understand existing phenomena in their natural settings. Given that this research did not seek to alter the variables or create experimental conditions but rather to analyze them as they naturally occur, this design was optimal. It provided a robust framework for statistical analysis, allowing me to draw meaningful inferences about the correlations between the chosen economic indicators and budget allocations. This methodological approach is particularly effective in economic and financial research where controlling variables is impossible. Thus, correlational design not only aligned with the objectives of this study but also adhered to the ethical standards of research in educational settings (Creswell & Creswell, 2018; Trochim & Donnelly, 2008).

## Research Methods

In this section, I describe the specific research methods I used to apply correlational research design in this study. Specifically, I discuss the setting, sample, data collection, data analysis, and steps taken to ensure validity or trustworthiness.

### Setting

The establishment and expansion of community colleges can be seen as a response to the increasing demand for college access in the early 1900s (A. Cohen & Brawer, 2003). The percentage of individuals completing high school and enrolling in college significantly increased during this time (A. Cohen & Brawer, 2003; Pasadena City College, 2002). Educators advocated for the creation of separate institutions, known as junior colleges, to provide general education for young people and relieve the burden on universities (Pasadena City College, 2002). However, community colleges did not become fully integrated into the mainstream until the middle of the century. The structure of community colleges initially involved them being extensions of secondary schools (A. Cohen & Brawer, 2003). The establishment of community colleges allowed for the accommodation of students with lower levels of preparedness and the provision of ongoing educational activities for individuals of all age groups. However, community colleges were also seen as alternative institutions and only a small number of creative structures that allowed transfer to upper-division universities managed to endure (Boggs & McPhail, 2016). The expansion of community colleges aimed to increase access to higher education and provide training for occupations between artisan jobs and professional careers (Pasadena City College, 2002). The historical path of community colleges, evolving from junior colleges intended to alleviate university burdens to becoming pivotal educational institutions, sets the stage for understanding the robust network of the California Community College system today (A. Cohen & Brawer, 2003). Originating from educational reforms in the early 20th century, this system has expanded to become the largest higher education system in the United States (Pasadena City College, 2002). Community colleges continue to fulfill their original purpose by catering to a wide range of individuals, including those who have historically been neglected in higher education.

As a quantitative study, this research delved into the key elements that constitute the state funding stream for CCCs, evaluating these variables for their potential to predict future trends. The California Community College system, the largest higher education system in the United States, comprises 116 colleges organized into 73 districts. It serves a diverse student body, with over 2.1 million students enrolled annually as of 2019 (California Community College Chancellor's Office [CCCCO], 2023).

Additionally, the California Community College system serves a significant number of first-generation college students and students who are returning to education after a prolonged absence (CCCCO, 2023; A. Cohen & Brawer, 2003). These demographic statistics emphasize the system serves not only as an educational institution but also to access better employment prospects and economic progress for many individuals. Designed to meet the diverse needs of this student body, an extensive array of programs is provided, spanning from academic transfer courses to vocational training. This diversity is a key aspect in the discussion of educational policies and budget allocations in the community colleges, as it underscores the need for flexible and inclusive educational strategies that address the specific needs of all student groups (CCCCO, 2024).

These colleges offer various educational programs, including college transfer courses, basic skills education, workforce training programs and services, and opportunities for lifelong learning (A. Cohen & Brawer, 2003). Each college in the system presents a unique production of programs, course offerings, and services, reflecting the distinct demographic profiles of their student populations and the specific business and industry characteristics of the communities they serve (CCCCO, 2024). The funding model for these colleges is complex, involving state allocations, local property taxes, and student fees, which can vary significantly from one district to another depending on local economic conditions and community priorities (Barr & McClellan, 2017). Additionally, the colleges have changing degrees of reliance on state funding, with some colleges supplementing their income through grants, partnerships with local businesses, and philanthropic contributions (Barr & McClellan, 2017; A. Cohen & Brawer, 2003). This financial diversity is further complicated by the colleges' responsibilities to offer degree programs, continuing education, and remedial courses, which traditional funding streams may need to cover fully (Barr & McClellan, 2017).

Researchers can analyze the financial characteristics of community colleges by comparing single-college districts with multicollege districts in California. Single college districts, which manage just one college, often experience more straightforward financial planning and budgeting processes due to their concentrated governance structures. These districts can tailor their financial strategies to the local community's needs, allowing for more agile responses to financial challenges. However, they may also face limitations in resource pooling and may have less bargaining power when procuring services and materials, which could lead to higher costs per student (Barr & McClellan, 2017; Vasquez Heilig et al., 2014).

Conversely, multicollege districts oversee multiple colleges within a larger geographic or demographic area. These districts benefit from economies of scale, as they can centralize many administrative functions and spread certain fixed costs across several institutions, potentially reducing overall expenses. They often have a more substantial financial base and can leverage more significant financial resources and fundraising capabilities, enhancing their stability and capacity for expansion (Barr & McClellan, 2017). However, this setup can also lead to more complex financial management challenges, including the need for more sophisticated financial governance structures and the potential for disparities in funding allocation among the colleges within the district (Barr & McClellan, 2017; Vasquez Heilig et al., 2014).

These diverse financial setups in single- and multi-college districts highlight the varied financial landscapes across the California Community College system, each with its advantages and challenges. This complexity underscores the necessity for tailored financial management strategies that align with each district's specific characteristics and needs to ensure that they continue to fulfill their educational missions effectively.

Whereas the findings from this study hold the potential to inform financial management and decision-making processes in higher education institutions globally and across the United States, the specific focus of the research is on the California Community College system. The system represents a unique sample within the broader landscape of higher education, characterized by its vast size, diversity, and California's specific regulatory and economic environment. By concentrating on this system, the study leverages the detailed financial data and varied economic conditions inherent to the state, providing a rich, contextual understanding of the dynamics at play in community college financing. Though derived from California's specific context, these insights may offer valuable lessons and strategies applicable to other regions and systems, reflecting broader trends in educational finance. The detailed exploration of this setting sets the stage for a focused examination of the sample, transitioning smoothly into a discussion of the specific data and methodologies employed to investigate these financial characteristics in California's community colleges.

**Sample**

The California Community College system is notable not just for its size but also for its exceptionally diverse student population, which mirrors the demographic complexities of California itself. The student body includes a significant representation from traditionally underrepresented groups, making it a critical institution for promoting higher education accessibility. In 2019, the CCCCO reported Hispanic students made up the largest ethnic group, accounting for nearly 46% of the overall student population, subsequently, 24% White, 11% Asian, 6% African American, 4% multiethnic, 3% Filipino, less than 1% American Indian and Alaskan Native, less than 1% Pacific Islander, and 6% unknown made up the remaining racial and ethnic demographics. Furthermore, 54% of students were women, 44% were men, less than 1% were nonbinary. The gender of 2% of students was unknown. This diversity extended beyond ethnicity to include a wide range of ages and socioeconomic backgrounds, with a substantial portion of students working either part- or full-time while attending classes (CCCCO, 2024). These figures highlight the increasing diversity of the community college population in California, reflecting the broader demographic direction toward a more diverse society. Concerning age distribution, data showed a substantial representation of adult learners: 29% of students were aged 30 or older. The 20–24 age group constituted approximately 23% of the total student population. In contrast, those aged 19 or younger made up 34%, reflecting a significant presence of younger students (CCCCO, 2024).

**Data Collection and Management**

The data collection and management section of this dissertation outlines the methodical process used to collect, evaluate, and protect the data necessary for studying the relationship between economic indicators and funding levels in CCCs. This method was essential for guaranteeing the study's results' accuracy, reliability, and integrity. To fully address the research questions, this section is divided into three main subsections: instrumentation, procedures, and data management.

*Instrumentation*

This study collected and used databases consisting of annual historical observations of state revenues and expenditure levels, variables representing California tax revenues, consist of corporation, sales, and local property taxes. The study also obtained data of the state's historic general fund disbursement levels to analyze relationships between expenditure levels and community college funding and examine predictive ability.

Moreover, this study needed to obtain comprehensive economic indicators that may influence state budget allocations to community colleges, including GDP growth rates, unemployment figures, and other relevant macroeconomic variables. Sources for these data sets included U.S. Census Bureau, the California State Department of Finance, the California Postsecondary Education Commission's Fiscal Profiles publication, and the California Legislative Analyst's Office. Such data were publicly available via the Internet.

The U.S. Census Bureau was a comprehensive source for this study, providing critical economic indicators necessary for examining advanced budgeting techniques in CCCs (Reddick, 2012). This federal agency collects extensive data on various economic variables, including unemployment rates and GDP growth, which were integral to understanding the economic environment impacting community college funding (Mansoor, 2021). The rich data sets available from the U.S. Census Bureau allowed for a detailed analysis of how broader economic trends correlate with changes in education funding. By accessing this reliable source, the study could incorporate precise, up-to-date economic statistics, enabling a robust analysis of the relationships between these economic indicators and the financial health of community colleges across California.

The California State Department of Finance is a government agency that offers extensive financial information, such as accurate budget data and economic predictions, for the state. This department releases yearly budget summaries that provide a detailed breakdown of the funds allocated to several sectors, including higher education (Roth, 2004). These resources were essential for comprehending the state's budgetary priorities and the fiscal context in which community colleges operate. They provided instructive insights into the government's funding methods and offer a macroeconomic perspective on the allocation of resources to education (Hovey, 2012; Roth, 2004).

The Fiscal Profiles publication by the California Postsecondary Education Commission provides a yearly summary of financial information pertaining to all sectors of California's higher education system, including community colleges (Commission for Postsecondary Education, 2023; Hovey, 2012). This publication was a vital resource for historical financial data and trend analysis, presenting information on revenue sources, expenditure patterns, and financial comparisons among institutions. The level of detail in the data enabled an examination of how funding is distributed and used in community colleges, facilitating an analysis of how budget allocations affect educational services and program offerings.

Lastly, the California Legislative Analyst's Office provides independent analyses of the state's budget, including assessments of financial proposals and their implications for public programs. The office's reports and forecasts were essential for understanding legislative changes, economic conditions, and their impacts on higher education funding (Hovey, 2012). Their analyses contributed to the contextualization of financial data within the wider framework of state economic policies and initiatives, offering an important perspective for any fiscal analysis pertaining to public institutions (e.g., community colleges). The combination of these sources provided an accurate dataset that allowed for a detailed analysis of the financial situations of CCCs. This dataset offered a full understanding of how economic and policy factors impact educational funding, with extensive and wide-ranging information.

### *Procedures*

The design of the procedures for data collection in this study ensured a thorough and systematic examination of the correlation between economic indicators and funding levels in CCCs. This process began with identifying and selecting relevant economic indicators (e.g., unemployment rates, GDP growth) which are widely recognized as significant predictors of public funding levels for higher education institutions (Hovey, 2012; Popesko et al., 2016). The study commenced with aggregating budget allocation data for all CCCs over the last 30 years. The initial data collection phase started withing the first 2 months of the project's initiation. During this period, efforts focused on gathering financial data and ensuring completeness and accuracy.

Following the aggregation of budget allocation data for CCCs, the next critical step involved collecting economic data from federal resources, primarily focusing on the U.S. Census Bureau website. This phase was dedicated to obtaining comprehensive economic indicators that may influence state budget allocations to community colleges, including GDP growth rates, unemployment figures, and other relevant macroeconomic variables over the last 30 years. The economic data collection process was scheduled to commence immediately following the completion of the budget data compilation, extending over the first 2 months. This timeline allowed for thorough research and data retrieval, ensuring all relevant economic indicators were accurately captured for the study's analysis phase.

### Data Management

Given the study's reliance on publicly available data, the data management strategy emphasized accuracy, reliability, and methodical organization rather than security concerns. The study employed Microsoft Excel spreadsheet software and SPSS statistical analysis software to efficiently organize and analyze these data. The study selected these tools due to their robust data handling, analysis capabilities, and widespread use in academic research, which enabled a structured approach to examining the dataset. Prior to analysis, I created and stored an initial backup of the raw data on an external hard drive and a cloud storage service, ensuring a pristine copy was preserved. The study similarly backed up the raw data and the results post analysis to prevent data loss and promote transparency in the research process. The methodical approach to data management grounded the study's findings in a thorough and accountable analysis of the economic factors influencing community college funding in California.

### Data Analysis and Interpretation

This section discusses the methodologies and statistical techniques I employed to scrutinize the relationships between economic indicators, fiscal policies, and the funding levels of CCCs. Using SPSS as the primary tool for statistical analysis, this section outlines a comprehensive approach to managing and interpreting the collected data. By integrating a variety of statistical tests, from exploratory data analysis and linear regression to the assessment of multicollinearity and variance, this analysis sought to establish a clear, empirical foundation for the study's hypotheses. In addition, this section explains the precise functions of these analyses in confirming the accuracy of the data and maintaining the reliability of the findings, consequently preparing the groundwork for well-informed decisions that could improve the strategic financial planning and operational efficiency of educational institutions in the state.

## Data Analysis

This study used the statistical analysis software SPSS to analyze extant data sets and conduct statistical tests. I constructed different linear regression analysis using each independent variable with potential predictive ability. In this study, the dependent variable was the funding levels for CCCs, which restated the total amount of financial resources allocated to these institutions annually. The independent variables included economic indicators (e.g., unemployment rates, GDP growth) and fiscal policies that impact income, corporations, sales, and local property taxes. Linear regression analysis also evaluated the predictive ability of the independent variables (Hovey, 2012; Leech et al., 2008). Multiple correlation coefficient ($R$) measured the strength of the linear relationship between several independent variables (e.g., economic indicators, fiscal policies) and one dependent variable (i.e., General Fund Tax Revenue [GFTR] for community colleges). It ranges from $-1$ to $1$, where values closer to 1 or $-1$ indicate a strong positive or negative linear relationship, respectively, and values near 0 indicate a weak relationship (Leech et al., 2008). A higher $R$ value suggested the combination of independent variables effectively captures the trends in GFTR revenue (Heck et al., 2010; Leech et al., 2008; Muijs, 2011).

Multiple regression analysis is a statistical technique to understand the relationship between one dependent variable and two or more independent variables (Heck et al., 2010). By incorporating multiple predictors, this study evaluated the simultaneous effect of various factors on the dependent outcome, adjusting for the influence of other variables in the model. This approach provided a more comprehensive analysis of the factors that significantly impacted the dependent variable, helping to isolate individual effects while controlling others (Leech et al., 2008). I reviewed the regression results to evaluate the significance of each independent variable's standardized regression coefficient, with each coefficient's *t* value judged to be significant at a *p* value < .05. If the independent variables exhibiting standardized coefficients that was not statistically significant, this study eliminated and the new model tested again for predictive ability, with the same tests used for the initial multiple regression (Heck et al.,

2010; Muijs, 2011). This time, I performed a test to determine the extent of correlation between the new set of independent variables and the residual.

### Procedures to Ensure Validity

Initially, this study analyzed the state's 30-year historical economic data to verify a normal

distribution's assumptions and identify possible outliers. Then, I used the descriptive analysis tool in

SPSS to identify the minimum and maximum values, the mean, the standard deviation, and the skewness

of each variable's data set, as Leech et al. (2008) recommended. After that, the study provided a scatter

plot diagram for linearity verification with SPSS output (Leech et al., 2008). The linearity between

independent and dependent variables is basic in statistical modeling, especially linear regression analysis

(Heck et al., 2010).

I mapped a correlation matrix to identify potential problems with multicollinearity among the independent variables. Leech et al. (2008) stated multicollinearity among independent variables is a situation in statistics where two or more independent variables in a regression model are highly correlated. High correlation means one variable can be linearly predicted from the others with a substantial degree of accuracy. There are some considerations regarding multicollinearity. It can make the estimates of the coefficients for the affected variables unreliable and unstable (Leech et al., 2008).

Small changes in the data or the model can lead to significant changes in the coefficient estimates. Then,

because the independent variables are closely related, it becomes difficult to determine which variable contributes to the dependent variable's variance. This collinearity complicates the interpretation of the model's coefficients because each predictor's individual effect on the dependent variable cannot be uniquely determined. Moreover, multicollinearity can inflate the standard errors of the coefficients. High standard errors can lead to coefficients that are not statistically significant, even if there is a strong underlying effect (Makridakis et al., 1998).

Leech et al. (2008) provided solutions, including removing, combining, and regularization variables. Removing variables is a common approach to resolving multicollinearity by removing one or more correlated independent variables from the model. Combining variables is another approach to combining the correlated variables into a single predictor through principal component analysis or factor analysis. Regularization refers to techniques such as Ridge or Lasso regression that can also be used. These methods add a penalty to the regression model that reduces the effect of collinear variables (Leech et al., 2008; Muijs, 2011).

This study also analyzed variance (ANOVA) and the results of the Levene test to check the violation of homogeneity of variances, which is a condition in statistical analyses where all groups or variables have equal variances across the range of predicted values (Leech et al., 2008). The violation of homogeneity of variances, often called heteroscedasticity, occurs in the context of statistical modeling, particularly in regression analysis. Homogeneity of variances is an assumption underlying many statistical tests, including ANOVA and linear regression, which states that the variances of the residuals (the differences between the observed values and the values predicted by the model) are constant across all levels of the independent variables (Heck et al., 2010).

Critical aspects of violating variances' homogeneity include estimates' efficiency and statistical inference. Although heteroscedasticity does not bias the regression coefficients themselves, it makes the estimates less precise. Variance estimates may become biased, which affects the calculation of confidence intervals and the significance of tests for the parameters (Leech et al., 2008). The standard errors associated with the regression coefficients may be underestimated or overestimated, leading to unreliable hypothesis tests (i.e., Type I and Type II errors). This misestimation can lead to incorrect conclusions about the significance of predictor variables (Heck et al., 2010; Leech et al., 2008).

Graphical residual plots and statistical tests allow the detection of heteroscedasticity. The first standard method to detect heteroscedasticity is through plotting the residuals against the predicted values or one of the independent variables. A pattern in the spread of residuals (e.g., a fan or cone shape) suggests heteroscedasticity. Several tests (e.g., Breusch-Pagan, White's, Goldfeld-Quandt) are specifically designed to detect heteroscedasticity in a regression model formally (Leech et al., 2008).

Remedies include transforming variables, weighted least squares, and robust standard errors. Transforming variables refers to applying transformations such as taking the logarithm, square root, or reciprocal of dependent and independent variables, which can help stabilize the variance. Weighted least squares involve giving each data point a weight that is inversely proportional to the variance of its residuals, compensating for heteroscedasticity. Robust standard errors can adjust the estimates to account for the heterogeneity in the residuals, thus providing more reliable statistical inference (Heck et al., 2010; Leech et al., 2008).

### Role of the Researcher

In quantitative research, the researcher's role is critical to ensuring the study's integrity and objectivity. This critical role of the researcher involves meticulous planning and execution at all stages of the research process, from the selection of instruments to the interpretation of data (Creswell & Creswell, 2018). I have carefully selected instruments in this study that are widely recognized and validated within the field of educational finance to ensure the reliability and relevance of the collected data to the research questions. I selected the instruments for collecting economic indicators (e.g., unemployment rates, GDP growth) as well as data on community college funding, due to their established validity and widespread use in academic and policy-making settings (Barr & McClellan, 2017).

Furthermore, I employed strategies to minimize bias and validate interpretations of the data. This included using the appropriate statistical methods for data and research questions (e.g., multiple regression analysis) to ensure that the results about the strong links between economic indicators and community college funding are not just a result of chance. I also remained open to alternative interpretations of the data, acknowledging the potential for confounding variables that could influence the results. To address the potential biases in the results, the study included checks for multicollinearity and employs control variables where necessary to isolate the effects of the primary independent variables. Additionally, I carefully constructed data collection methods to avoid leading questions that could skew the results. By maintaining a rigorous and reflective approach throughout the research process, I aimed to provide findings that were both statistically significant and substantively meaningful, contributing valuable insights into the financial dynamics affecting community colleges.

## Chapter Summary

Chapter 3 of this dissertation outlined the methodological framework to examine the alignment of advanced budgeting techniques with the strategic missions and budgetary efficiency in CCCs. This chapter systematically discussed the quantitative techniques in this research to address resource management and budgetary responsibility. Moreover, it comprehensively explained the methodological approach, which used correlational research designs and other statistical analyses to understand the complex connections between economic variables and funding levels. It discussed these institutions' sophisticated financial management techniques, emphasizing the importance of economic indicators and fiscal policies in determining funding levels. In addition, this summary provides an explanation of the chapter's content by highlighting the importance of the study in addressing the problem statement and its potential impact on future research, which this study further explores in Chapter 4. This dissertation examined how advanced budgeting techniques can help promote CCCs' strategic missions and objectives. The problem investigated was the possible misalignment and its influence on the colleges' capacity to handle resources and uphold budgetary effectiveness efficiently. This issue is important because it directly relates to the fundamental operations of these institutions in delivering accessible, high-quality education while maintaining financial responsibility (Hovey, 2012).

This chapter begun by discussing quantitative research methodology, which relies on empirical data and allows the investigation of the connections between economic indicators (e.g., GDP growth, unemployment rates, funding levels) inside community colleges. The study used a correlational design, which involves observing the natural relationship between variables rather than manipulating them (Creswell & Creswell, 2018; Leavy, 2023). The goal was to identify patterns that might assist community college leaders in anticipating future budget fluctuations and informed resource allocations. Quantitative tools in this work were especially relevant because they could effectively manage massive data sets and generate unbiased, widely applicable findings (Creswell & Creswell, 2018). Ensuring unbiased and widely applicable findings is of utmost importance in a context like the California Community College system, where financial operations are intricate and impacted by multiple economic factors (Hovey, 2012). The research design carefully incorporated the different elements of the study (e.g., sample selection, data collection, statistical testing) to achieve a unified approach to comprehending these financial dynamics.

A thorough analysis of the sample highlighted the scope of data, comprising diverse economic indicators and fiscal policies. The data collection process was robust, tapping into reliable sources like the U.S. Census Bureau and the California State Department of Finance. This reliance on data from sources like the U.S. Census Bureau and the California State Department of Finance ensured the economic variables analyzed accurately reflected the real-world scenarios affecting community colleges. Statistical analyses (e.g., linear regression, multiple correlation coefficients) formed the core of the study's methodology. These techniques assessed the strength and direction of relationships between variables (Cornell, 2020; Creswell & Creswell, 2018), providing insights into how economic trends might influence future funding decisions. This statistical approach strengthened the dependability of the results and guaranteed the inferences made were grounded in empirical data. In addition, the chapter discussed potential limitations of quantitative research, such as its sometimes-superficial depth regarding complex human behaviors and the rigidity that might overlook novel insights. However, the planned comprehensive data analysis aimed to mitigate these weaknesses by thoroughly examining the involved variables. Additionally, to ensure the validity and trustworthiness of the analysis, this study used various techniques, including verification of distribution of assumptions, scatter plot diagrams, correlation matrix, Handling multicollinearity, analysis of variance, graphical residual plot and test (Heck et al., 2010; Leech et al., 2008; Muijs, 2011).

The methodological foundation established in this chapter prepared for the presentation of empirical results in Chapter 4. I anticipated the findings would demonstrate the practical consequences of the connections between economic indicators and funding levels. By correlating these variables, the study offered significant insights into how CCCs can optimize their budgetary strategies to better align with their educational missions (Barr & McClellan, 2017; Hovey, 2012). The forthcoming chapter explores the specific results of the analyses organized around the research questions. The discussion explores the implications of these findings for financial management in higher education, offering practical ideas that have the potential to impact policy and strategic decision making in community colleges throughout California.

# CHAPTER 4

# FINDINGS

Chapter 4 presents findings from this quantitative study, examining the interaction between advanced budget techniques, budget policies, and funding levels in California Community Colleges (CCCs). From the theoretical and methodological foundations discussed in the previous chapters, Chapter 4 answers the research questions through a thorough analysis of the data gathered. I discuss the findings in relation to the study's research questions, with each section providing an extensive description of the data, statistical results, and my relevant observations. The chapter maintains a strict empirical focus without drawing inferences and conclusions, which are reserved in Chapter 5. By linking the findings to the purpose and extent of the study, Chapter 4 provides an organized and objective foundation to understand budget practices and their implications to community colleges.

## First Research Question

I investigated Research Question 1, "What is the relationship between advanced budgeting techniques' economic indicators (e.g., unemployment rates, gross domestic product [GDP] growth) and funding levels for CCCs and students?" To answer this question, I developed and tested six hypotheses, each addressing a specific economic indicator's relationship to funding levels. These hypotheses allowed me to examine how broader economic conditions influenced the financial resources allocated to CCCs. The six hypotheses I tested are described in the following paragraphs.

### H1a

H1a stated there was a statistically significant relationship between the funding levels for CCCs and GDP growth rate in the United States.

I conducted a simple linear regression analysis to examine the relationship between California Community College funding levels and GDP growth rate in the United States. In this analysis, I used GDP change as the independent variable and community college funding level as the dependent variable. Before running the regression analysis, I assessed the assumptions of normality and linearity.

To evaluate normality, I examined skewness and kurtosis values. The results showed GDP change had a skewness of –0.328 and a kurtosis of 2.241, whereas community college funding levels had a skewness of 0.529 and a kurtosis of –0.575. These values indicated, although the data distribution was not perfectly normal, it remained within an acceptable range for regression analysis. I also tested linearity using a scatter plot, which revealed a weak positive trend between GDP change and community college funding levels.

The regression analysis results indicated a weak and nonsignificant relationship between GDP growth rate and community college funding levels. The scatter plot and fitted regression line showed a near-horizontal trend, and I found the $R^2$ value was 0.006, meaning GDP change explained only 0.6% of the variance in funding levels. Additionally, I calculated Pearson's correlation coefficient, which was 0.078, with a significance level of 0.684. These results indicated the relationship was not statistically significant. Therefore, I did not find support for H1a, which stated there was a statistically significant relationship between California Community College funding levels and GDP growth rate. These findings suggested other economic or policy-related factors likely played a more substantial role in determining funding allocations for community colleges in California.

**H1b**

H1b indicated there was a statistically significant relationship between the funding levels for CCCs and GDP growth in dollar amounts in the United States.

I conducted a simple linear regression analysis to examine the relationship between California Community College funding levels and GDP growth in dollar amounts in the United States. Figure 2 shows the scatterplot for the correlation between U.S. GDP in billions of dollars and CCC funding levels from 1994 to 2023. The strong linear relationship ($R^2 = 0.928$) indicated increases in national GDP are closely associated with increases in CCC funding levels.

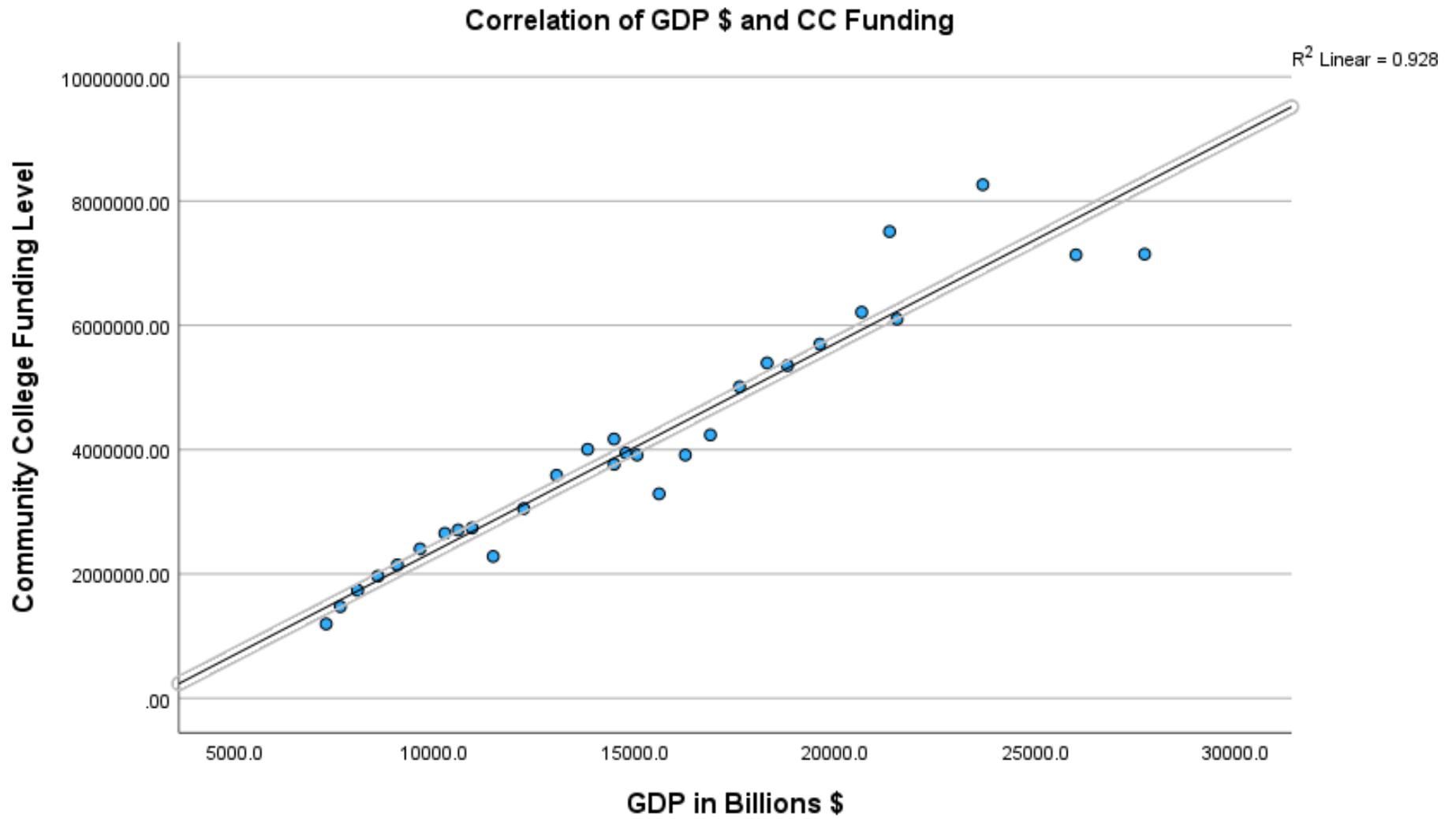

*Figure 2.* Scatterplot showing the correlation between U.S. GDP and California Community College funding, 1994–2023.

In this analysis, I used GDP in billions of dollars as the independent variable and community college funding level as the dependent variable. Before running the regression analysis, I assessed the assumptions of normality and linearity. The skewness and kurtosis values for GDP in billions of dollars were 0.517 and –0.411, respectively, whereas community college funding levels had a skewness of 0.529 and a kurtosis of –0.575. These values indicated the data distribution was within an acceptable range for regression analysis. I also tested linearity using a scatter plot, which revealed a strong positive trend between GDP in billions of dollars and community college funding levels.

The regression analysis results indicated a strong and statistically significant relationship between GDP in dollar amounts and community college funding levels. The scatter plot and fitted regression line showed a clear upward trend, and I found that the $R^2$ value was 0.928, meaning GDP in billions of dollars explained 92.8% of the variance in funding levels. Additionally, I calculated Pearson's correlation coefficient, which was 0.964, with a significance level of $p < .001$, confirming the relationship was statistically significant. Therefore, I found strong support for H1b, which stated there was a statistically significant relationship between California Community College funding levels and GDP growth in dollar amounts. These findings suggested, as the GDP in the United States increases, funding for CCCs also rises in a highly correlated manner, highlighting the strong dependence of community college funding on overall economic growth.

## H1c

H1c stated there was a statistically significant relationship between the funding levels for CCCs and the inflation rate in the United States.

I conducted a simple linear regression analysis to examine the relationship between California Community College funding levels and the inflation rate in the United States. In this analysis, I used the inflation rate as the independent variable and community college funding level as the dependent variable. Before running the regression analysis, I assessed the assumptions of normality and linearity. The skewness and kurtosis values for the inflation rate were 1.510 and 5.583, respectively, indicating a moderate deviation from normality. Community college funding levels had a skewness of 0.529 and a kurtosis of –0.575, which remained within an acceptable range for regression analysis. I also tested linearity using a scatter plot, which showed a weak positive trend between the inflation rate and community college funding levels.

The regression analysis results indicated a weak and nonsignificant relationship between the inflation rate and community college funding levels. The scatter plot and fitted regression line revealed a slight upward trend, and I found the $R^2$ value was 0.053, meaning the inflation rate explained only 5.3% of the variance in funding levels. Additionally, I calculated Pearson's correlation coefficient, which was 0.231, with a significance level of 0.220. These results indicated the relationship was not statistically significant. Therefore, I did not find support for H1c, which stated there was a statistically significant relationship between California Community College funding levels and the inflation rate. These findings suggested, although inflation may have some influence on funding decisions, other economic or policy-driven factors likely play a more significant role in determining funding allocations for community colleges in California.

## H1d

H1d stated there was a statistically significant relationship between the funding levels for CCCs and the Consumer Price Index (CPI) in dollar amounts in the United States.

I conducted a simple linear regression analysis to examine the relationship between California Community College funding levels and the CPI in dollar amounts in the United States. In this analysis, I used CPI in dollars as the independent variable and community college funding level as the dependent variable. Before running the regression analysis, I assessed the assumptions of normality and linearity. The skewness and kurtosis values for CPI in dollars were 0.295 and –0.639, respectively, whereas community college funding levels had a skewness of 0.529 and a kurtosis of –0.575. These values indicated the data distribution was within an acceptable range for regression analysis. I also tested linearity using a scatter plot, which revealed a strong positive trend between CPI in dollars and community college funding levels.

The regression analysis results indicated a strong and statistically significant relationship between CPI in dollar amounts and community college funding levels, as shown in Figure 3. The scatter plot and fitted regression line showed a clear upward trend, and I found the $R^2$ value was 0.900, meaning CPI in dollars explained 90.0% of the variance in funding levels. Additionally, I calculated Pearson's correlation coefficient, which was 0.949, with a significance level of $p < .001$, confirming the relationship was statistically significant. Therefore, I found strong support for H1d, which stated there was a statistically significant relationship between California Community College funding levels and CPI in dollar amounts. These findings suggested, as the cost of goods and services measured by CPI increases, funding for CCCs also rises in a highly correlated manner, highlighting the strong association between inflationary trends and educational funding.

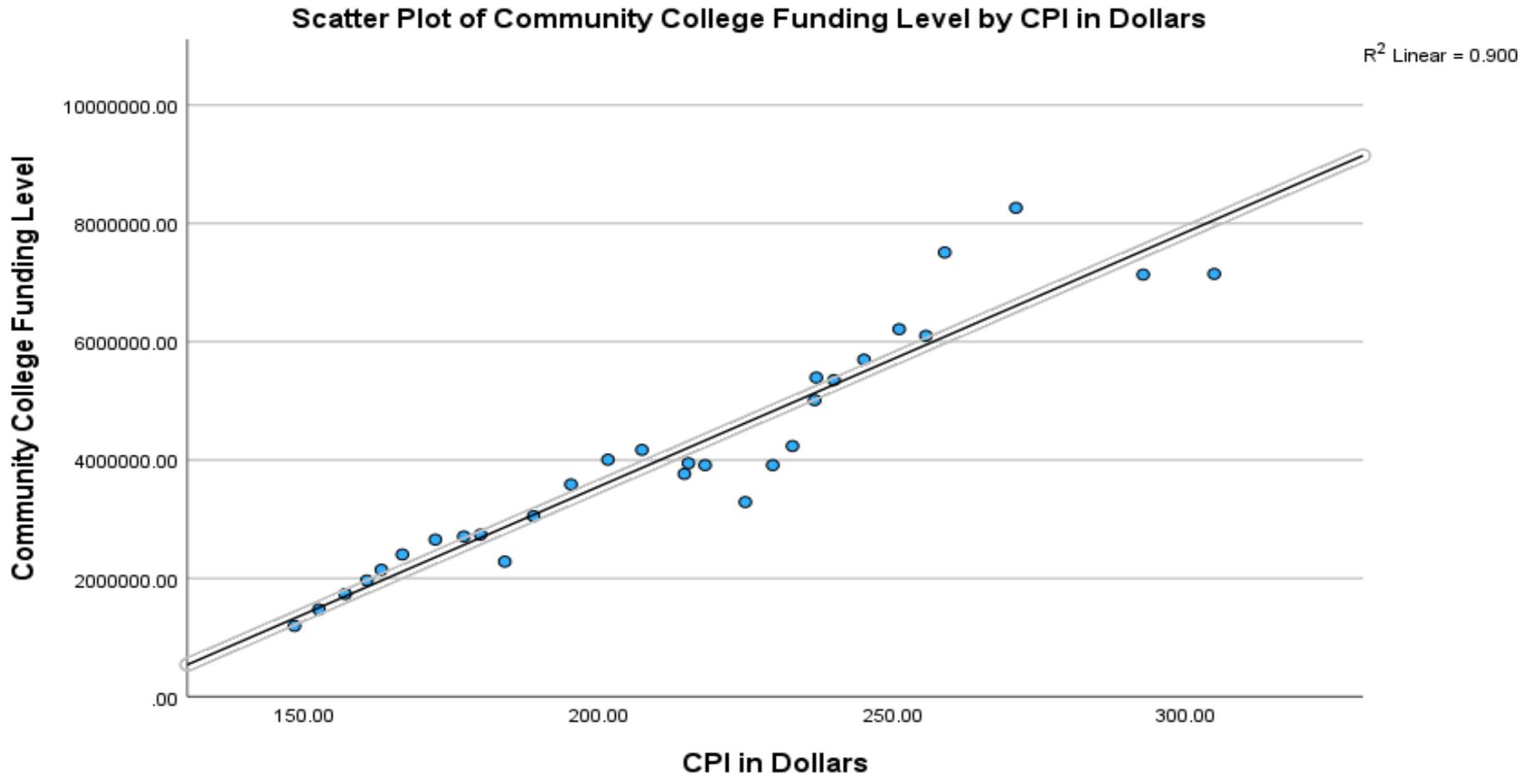

*Figure 3.* Scatterplot showing the relationship between CPI and California Community College funding, 1994–2023 (*R²* = .90).

**H1e**

H1e claimed there was a statistically significant relationship between the funding levels for CCCs and the unemployment rate in the United States.

I conducted a simple linear regression analysis to examine the relationship between California Community College funding levels and the unemployment rate in the United States. In this analysis, I used the U.S. unemployment rate as the independent variable and community college funding level as the dependent variable. Before running the regression analysis, I assessed the assumptions of normality and linearity. The skewness and kurtosis values for the U.S. unemployment rate were 1.049 and 0.284, respectively, indicating mild deviation from normality but remaining within an acceptable range for regression analysis. Community college funding levels had a skewness of 0.529 and a kurtosis of –0.575, confirming the data distribution was suitable for regression analysis. I also tested linearity using a scatter plot, which showed a weak negative trend between the U.S. unemployment rate and community college funding levels.

The regression analysis results indicated a weak and nonsignificant relationship between the U.S. unemployment rate and community college funding levels. The scatter plot and fitted regression line displayed a slight downward trend, and I found the $R^2$ value was 0.016, meaning the U.S. unemployment rate explained only 1.6% of the variance in funding levels. Additionally, I calculated Pearson's correlation coefficient, which was –0.126, with a significance level of 0.507. These results indicated the relationship was not statistically significant. Therefore, I did not find support for H1e, which stated there was a statistically significant relationship between California Community College funding levels and the U.S. unemployment rate. These findings suggested national unemployment levels do not play a direct or substantial role in determining funding allocations for community colleges in California, implying other economic or policy-related factors likely have a greater impact.

**H1f**

H1f stated there was a statistically significant relationship between the funding levels for CCCs and the unemployment rate in California.

I conducted a simple linear regression analysis to examine the relationship between California Community College funding levels and the unemployment rate in California. In this analysis, I used the California unemployment rate as the independent variable and community college funding level as the dependent variable. Before running the regression analysis, I assessed the assumptions of normality and linearity. The skewness and kurtosis values for the California unemployment rate were 0.971 and 0.054, respectively, indicating mild deviation from normality but remaining within an acceptable range for regression analysis. Community college funding levels had a skewness of 0.529 and a kurtosis of –0.575, confirming the data distribution was suitable for regression analysis. I also tested linearity using a scatter plot, which showed a weak negative trend between the California unemployment rate and community college funding levels.

The regression analysis results indicated a weak and nonsignificant relationship between the California unemployment rate and community college funding levels. The scatter plot and fitted regression line displayed a slight downward trend, and I found the $R^2$ value was 0.026, meaning the California unemployment rate explained only 2.6% of the variance in funding levels. Additionally, I calculated Pearson's correlation coefficient, which was –0.162, with a significance level of 0.392. These results indicated the relationship was not statistically significant. Therefore, I did not find support for H1f, which stated there was a statistically significant relationship between California Community College funding levels and the California unemployment rate. These findings suggested fluctuations in state-level unemployment do not have a direct or substantial impact on community college funding allocations, implying other economic or policy-driven factors play a more significant role in determining funding levels.

## Second Research Question

I investigated Research Question 2, "What is the relationship between fiscal policies (i.e., policies impacting income, corporate, sales, and local property taxes) and funding levels for CCCs and students?" To respond to this question, I created and tested four hypotheses, which covered multiple aspects of California's fiscal policies and their impacts on funding levels. The four hypotheses I tested enabled me to establish how state-level tax revenues impacted the funding to CCCs. The four hypotheses I analyzed are included in the following paragraphs.

### H2a

H2a indicated there is a statistically significant relationship between the funding levels for CCCs and California corporate tax income.

I conducted a simple linear regression analysis to examine the relationship between California Community College funding levels and California corporate tax income. In this analysis, I used California corporate tax income as the independent variable and community college funding level as the dependent variable. I tested the assumptions of normality and linearity before conducting the regression analysis. The skewness and kurtosis values for California corporate tax income were 2.925 and 8.619, respectively, indicating a higher deviation from normality. However, community college funding levels had a skewness of 0.529 and a kurtosis of –0.575, remaining within an acceptable range for regression analysis. I also tested linearity using a scatter plot, which showed a moderate positive trend between California corporate tax income and community college funding levels.

The regression analysis results indicated a statistically significant relationship between California corporate tax income and community college funding levels. The regression line and scatterplot had a clear positive slope. I found that the $R^2$ value was 0.503, meaning California corporate tax income explained 50.3% of the variance in funding levels. Moreover, I calculated Pearson's correlation coefficient, which was 0.710, with a significance level of $p < .001$, confirming the relationship was statistically significant. Therefore, I found support for H2a, which stated there was a statistically significant relationship between California Community College funding levels and California corporate tax income. These findings demonstrated corporate tax revenue is a key determinant of community college funding allocations in California, and they highlight the importance economic activity and business taxation have in financing higher education.

## H2b

H2b stated there is a statistically significant relationship between the funding levels for CCCs and California personal tax income.

I conducted a simple linear regression analysis to examine the relationship between California Community College funding levels and California personal income tax revenue. In this analysis, I used California personal income tax as the independent variable and community college funding level as the dependent variable. Before running the regression analysis, I assessed the assumptions of normality and linearity. The skewness and kurtosis values for California personal income tax were 0.857 and 0.006, respectively, indicating the data distribution was within an acceptable range for regression analysis. Community college funding levels had a skewness of 0.529 and a kurtosis of –0.575, confirming suitability for regression analysis. I also tested linearity using a scatter plot, which showed a strong positive trend between California personal income tax and community college funding levels.

The regression analysis results indicated a strong and statistically significant relationship between California personal income tax revenue and community college funding levels. The scatter plot and fitted regression line displayed a clear upward trend, and I found that the $R^2$ value was 0.889, meaning California personal income tax explained 88.9% of the variance in funding levels. Additionally, I calculated Pearson's correlation coefficient, which was 0.943, with a significance level of $p < .001$, confirming the relationship was statistically significant. Therefore, I found strong support for H2b, which stated there was a statistically significant relationship between California Community College funding levels and California personal income tax revenue. These findings suggested personal income tax revenue plays a crucial role in determining funding allocations for community colleges in California, emphasizing the reliance of higher education funding on individual tax contributions.

**H2c**

H2c said there is a statistically significant relationship between the funding levels for CCCs and California sales tax revenue.

I conducted a simple linear regression analysis to examine the relationship between California Community College funding levels and California sales tax revenue. In this analysis, I used California retail sales and used tax as the independent variable and community college funding level as the dependent variable. Before running the regression analysis, I assessed the assumptions of normality and linearity. The skewness and kurtosis values for retail sales and use tax were 0.150 and 0.011, respectively, indicating the data distribution was within an acceptable range for regression analysis. Community college funding levels had a skewness of 0.529 and a kurtosis of –0.575, confirming suitability for regression analysis. I also tested linearity using a scatter plot, which showed a moderate positive trend between California sales tax revenue and community college funding levels.

The regression analysis results indicated a statistically significant relationship between California sales tax revenue and community college funding levels. The scatter plot and fitted regression line displayed a clear upward trend, and I found that the $R^2$ value was 0.718, meaning California sales tax revenue explained 71.8% of the variance in funding levels. Additionally, I calculated Pearson's correlation coefficient, which was 0.718, with a significance level of $p < .001$, confirming the relationship was statistically significant. I confirmed, therefore, support for H2c, which held that there was a statistically significant relationship between California sales tax collection and California community college funding levels. These results showed sales tax dollars directly affect how funds are distributed. They also highlighted the role of consumer spending and economic activity in supporting higher education.

**H2d**

H2d stated there is a statistically significant relationship between the funding levels for CCCs and total tax income in California.

I conducted a simple linear regression analysis to examine the relationship between California Community College funding levels and total tax income in California. In this analysis, I used total tax income (i.e., corporate tax, personal income tax, and sales tax) as the independent variable and community college funding level as the dependent variable. Before running the regression analysis, I assessed the assumptions of normality and linearity. The skewness and kurtosis values for total tax income were within an acceptable range for regression analysis, with personal income tax, corporate tax, and sales tax individually displaying normality characteristics. Community college funding levels had a skewness of 0.529 and a kurtosis of –0.575, confirming suitability for regression analysis. I also tested linearity using a scatter plot, which showed a strong positive trend between total tax income and community college funding levels.

The regression analysis results indicated a strong and statistically significant relationship between total tax income and community college funding levels. The scatter plot and fitted regression line displayed a clear upward trend, and I found that the $R^2$ value was 0.889, meaning total tax income explained 88.9% of the variance in funding levels. Additionally, I calculated Pearson's correlation coefficient, which was 0.943, with a significance level of $p < .001$, confirming the relationship was statistically significant. Therefore, I found strong support for H2d, which stated there was a statistically significant relationship between California Community College funding levels and total tax income in California. These findings suggested state tax revenues, including personal income tax, corporate tax, and sales tax, played a crucial role in determining funding allocations for community colleges in California, reinforcing the connection between tax collection and higher education funding stability. Through these hypotheses, I explored the connections between various tax revenues and community college funding, providing a comprehensive analysis of fiscal policy impacts.

**Chapter Summary**

In this chapter, I analyzed the relationships between California Community College funding levels and various economic indicators, including GDP growth rate, GDP in dollar amounts, inflation rate, CPI, U.S. and California unemployment rates, and state tax revenues. The findings revealed mixed results, with some economic factors demonstrating significant associations with funding levels, although others did not. I found no statistically significant relationships between community college funding and GDP growth rate, U.S. and California unemployment rates, or the inflation rate, suggesting these macroeconomic variables do not directly influence funding allocations. However, I identified strong and statistically significant relationships between funding levels and GDP in dollar amounts, CPI, and state tax revenues, highlighting the critical role of overall economic growth and tax income in determining funding for community colleges.

Among the most notable findings, I found GDP in dollar amounts and CPI had strong positive correlations with community college funding levels, with GDP explaining 92.8% of the variance and CPI explaining 90.0%. I also found significant correlations between funding levels and corporate tax revenue, personal income tax, sales tax, and total tax revenue, with personal income tax showing the strongest relationship ($r = .943$, $p < .001$). These results underscored the importance of tax revenue as a primary determinant of community college funding in California. In contrast, unemployment rates and GDP percentage change appeared to have little effect, suggesting funding sources were more closely tied to direct economic performance and state budget policies rather than short-term labor market trends. These findings set the foundation for Chapter 5, where I discuss their implications, provide interpretations, and offer recommendations based on the results.

# CHAPTER 5

# DISCUSSION

This study examined the critical role budget forecasting and strategic planning have in assisting decision making by California community college leaders. The economic situation, state funding formula, and enrollment trends all contribute to the dynamic situation in higher education, and leaders employ data-driven forecasting to maintain fiscal sustainability and operational efficiency (Dougherty & Natow, 2015). These forecasting techniques can help institutions navigate times of uncertainty and make informed decisions to prepare them for the future. The problem addressed by this study was the increased sophistication in budget planning in community colleges, where uncertain funding and policy evolution present challenges to long-term budget stability. Without reliable forecasting systems and strategic planning, institutions are vulnerable to inefficiencies, misallocations, and potential disruption to initiatives promoting students' success (Dimitrijevska-Markoski et al., 2021).

The purpose of this quantitative study was to examine the effectiveness of budget forecasting models in supporting strategic decision making by community college leaders. By examining current forecasting practices and their congruence with institutional planning, the study attempted to establish best practices that result in more informed operational and budget decisions. By collecting and comparing budget data from all 116 community colleges in California, the study illuminated how budget forecasting enhances institutional flexibility, resource allocation, and long-term sustainability. Findings from the study added to the broader debate surrounding higher education finance by offering evidence-based recommendations to increase the resilience of community college finances.

This study was guided by two primary research questions aimed at examining the relationship between economic and fiscal factors and the funding levels for California Community Colleges (CCCs). The first research question explored the relationship between advanced budgeting techniques' economic indicators (i.e., including gross domestic product (GDP) growth rate, GDP in dollar amounts, inflation rate, Consumer Price Index (CPI), U.S. unemployment rate, and California unemployment rate) and funding levels for CCCs. The second research question examined the relationship between fiscal policies, including California corporate tax income, personal tax income, sales tax revenue, and total tax income, and the funding levels for CCCs. To address these research questions, I tested a series of hypotheses that quantified the strength and significance of these relationships.

I employed a quantitative research design grounded in correlational and regression analyses to answer these research questions. This design allowed me to explore the correlation between key economic and fiscal variables and funding to CCCs. I collected secondary data from publicly available financial reports, government databases, and institutional records spanning multiple years. I used Pearson's correlation analysis to measure the strength and direction of relationships between economic indicators and community college funding. I also used simple linear regression analyses to determine how much each independent variable, such as GDP in dollars, CPI, or tax revenues, accounted for funding level variation. This design allowed a rigorous analysis of finances and an evidence-based foundation to explore budget forecasting and fiscal planning in higher education.

## Interpretations

This section provides an interpretation of the findings from the study considering the research questions and hypotheses. The results are situated in broader literature on budget forecasting, financial management, and equitable resource distribution in higher education. The results shed light on economic and fiscal determinants that influence funding levels in CCCs, and which variables have high prediction ability. Understanding these budget trends is significant to promote efficiency in budgeting and to provide continuous support to (diversity, equity, and inclusion (DEI) initiatives that serve historically marginalized students.

Through an analysis of these findings, I can establish trends that can inform institutional decision making so that funding approaches are in keeping with California Community College's vision to provide high-quality, accessible education to all students, including first-generation, low-income, and underrepresented students. This analysis identifies potential implications for community college leaders, policymakers, and stakeholders to better allocate resources and budget in ways that prioritize equity, increase DEI initiatives, and foster an inclusive learning environment where all students have the resources they need to excel. An understanding of the economic forces that drive funding decisions can help institutions develop anticipatory strategies that protect key student support services, including DEI offices, disability resource centers, extended opportunity programs and services (EOPS), TRIO programs, and other equity-based initiatives, from budget uncertainty.

## Summary of Major Findings and Interpretations

The following sections present a summary of the study's key statistical findings and interpretations. These interpretations highlight which economic indicators are most closely related to community college funding trends in California.

### *GDP and Inflation Rate*

Hovey (2012) previously examined the correlation between funding levels of community colleges and economic indicators and found no correlation between GDP and inflation rate and funding levels of CCCs. The findings of the current study negate that result by establishing a strong, statistically significant correlation between GDP dollar growth and funding levels of community colleges. The current study also establishes that CPI, which is the most important component used to determine the inflation rate, is significantly related to funding level of CCCs. These results establish, as the economy expands, community college funding also increases, confirming the relevance of including macroeconomic trends in budget forecasting.

The findings of the current study establish economic and fiscal indicators that are significantly related to funding levels of community colleges, and there are others that have no relation. One of the most surprising results was the strong, statistically significant correlation between GDP dollar growth and funding levels of community colleges ($R^2 = 0.928$, $p < .001$), which established 92.8% of the variation in funding levels of community colleges was explained by variation in GDP dollar growth, which is an extremely strong correlation to predict. The $p$ value less than .001 established the correlation was statistically significant, implying a 99.9% chance the result was not by chance and that GDP dollar growth is an important driver of funding levels of community colleges. This result establishes that economic expansion as a whole directly impacts how much funding these institutions provide.

Similarly, the CPI in dollar values was highly positively related to funding levels ($R^2 = 0.900$, $p < .001$), and 90.0% of the variance in funding levels in community colleges was accounted for by variation in the CPI in dollar values, and there is a high prediction relationship. The $p$ value below .001 indicated a 99.9% probability this result was not by chance, and it was established that rising prices of goods and services were highly associated with rising funding allocations to community colleges. This confirms the theory that funding also rises due to pressure from inflation.

Hovey (2012) found no correlation between funding levels and the inflation rate, a result supported by the findings in this study. Similarly, I found the funding levels were not statistically influenced by the inflation rate, supporting the discovery that year-to-year change in the inflation rate will not have a direct influence on budget allocations to CCCs. One explanation for the finding is how the inflation rate is measured. The inflation rate is measured by the change in the CPI over time, and it is a relative change, not an absolute level. This conversion makes the raw data on inflation less stable and more susceptible to year-to-year fluctuations that may not indicate long-term trends in finances. As a result, the inflation rate may not be a good indicator of funding trends because it is measured by year-to-year percent change and not by the level of costs that influence budgeting.

To overcome this limitation, academics should accord more importance to CPI in dollar values rather than the inflation rate when they study economic relations. Unlike the inflation rate, CPI is the real price level of goods and services over time and is a more stable and cumulative indicator of economic conditions. This stability allows CPI to better capture the long-term economic pressures underlying funding decisions in higher education.

Rabovsky (2012) and Popesko et al. (2016) suggested GDP percentage changes may not accurately capture economic expansion effects on education funding. Furthermore, Hovey (2012) found no significant correlation between GDP and community college funding levels, but only based on GDP growth rate (i.e., percentage change in GDP over a period) and not GDP dollar values. Similarly, I also found no statistically significant correlation between GDP growth rate and community college funding levels ($R^2 = 0.006$, $p = 0.684$), indicating GDP percentage change over a short period is not significantly associated with the amount of funding allocated to CCCs. This finding is significant in highlighting a methodological limitation: GDP growth rate, as a percent measure, reflects the relative economic expansion or contraction but not the size of the economy, and could potentially obscure the available real financial resources to support higher education. To address this limitation, I employed GDP dollar values instead of GDP growth rate and found a strong and statistically significant correlation between GDP dollar values and community college funding levels.

### Unemployment

The findings of this study revealed no statistically significant correlation between funding levels to community colleges and the unemployment rate, in California and the United States. The findings for H1e (i.e., U.S. unemployment rate) showed an $R^2 = 0.016$ and a Pearson's correlation coefficient of –0.126 ($p = 0.507$), indicating only 1.6% of the variance in funding levels could be explained by changes in the U.S. unemployment rate. Similarly, the findings for H1f (California unemployment rate) showed an $R^2 = 0.026$ and a Pearson's correlation coefficient of –0.162 ($p = 0.392$), meaning only 2.6% of the variance in community college funding could be explained by state-level unemployment fluctuations. Because both results did not meet statistical significance, these findings suggested unemployment levels, whether at the national or state level, do not directly impact funding allocations to CCCs.

These results were supported by Hovey (2012), who also found no correlation between unemployment rates and community college funding levels. One possible explanation is that although unemployment rates fluctuate based on economic cycles, they do not directly determine state funding decisions for higher education. In theory, rising unemployment could increase community college enrollments as displaced workers seek job training, potentially leading to higher funding allocations.

However, in practice, state funding is more influenced by tax revenue trends, economic growth, and budgetary constraints rather than changes in employment levels. The findings in this study confirmed unemployment rates alone are weak predictors of community college funding trends, whereas fiscal policy factors (e.g., state tax revenues, GDP in dollar values) serve as more accurate indicators of funding allocations.

Additionally, the weak negative correlations observed in both cases (i.e., –0.126 for U.S. unemployment and –0.162 for California unemployment) suggest a slight inverse relationship, meaning as unemployment rises, funding tends to decline slightly. This trend could be because higher unemployment is often a sign of an economic downturn, which can reduce state revenues and lead to budget cuts rather than increased funding for community colleges. As a result, although workforce retraining is a core function of community colleges, funding does not appear to increase in response to rising unemployment levels. Instead, funding decisions are more likely to be based on broader economic and fiscal trends rather than labor market fluctuations alone.

### *Fiscal Policy Variables*

The second research question explored the connection between fiscal policy variables (i.e., corporate tax income, personal income tax, sales tax revenue, and total California tax income) and funding levels for CCCs. My findings revealed strong and statistically significant relationships between CCC funding and each of these fiscal variables.

For corporate tax income (H2a), the analysis showed a statistically significant relationship, with an $R^2$ value of 0.503 and a Pearson correlation coefficient of 0.710 ($p < .001$). This finding suggested corporate tax revenue explains approximately 50.3% of the variation in CCC funding. These results highlighted the critical role corporate tax level play in supporting community colleges' financial stability. However, Hovey (2012) did not find a statistically significant relationship between CCC funding and corporate tax revenue. This discrepancy may stem from differences in methodology, including variations in the timeframe and economic conditions analyzed. My study suggested corporate tax revenue plays a larger role than previously recognized, possibly due to shifts in corporate tax policies and economic trends over time.

My findings revealed an even more robust link with CCC funding for personal income tax revenue (H2b). Personal income tax revenue explains 88.9% of the variance in CCC funding levels according to the regression analysis, which revealed an $R^2$ value of 0.889 and a Pearson correlation coefficient of 0.943 ($p <.001$). This result implies CCCs mostly depend on income tax collections, which reflects the general trend of state funding policies strongly dependent on personal income. Hovey (2012) nevertheless discovered no appreciable correlation between personal income tax receipts and CCC financing. My results contrast with these findings, most likely because of changes in tax policies and California's increasing reliance on personal income tax as a main source of CCC funding.

For sales tax revenue (H2c), my study confirmed a statistically significant relationship, with an $R^2$ value of 0.718 and a Pearson correlation coefficient of 0.718 ($p < .001$). This result indicates 71.8% of the variance in CCC funding can be attributed to fluctuations in sales tax revenue, reinforcing the idea that economic activity and consumer spending significantly impact higher education funding. Hovey (2012) found no significant relationship between sales tax revenue and CCC funding. Based on my results, sales tax income seems to be becoming increasingly significant as a funding source maybe because of changes in state budget distribution and varied time frames applied.

Lastly, for total tax income (H2d), my results revealed a strong correlation, with an $R^2$ value of 0.889 and a Pearson correlation coefficient of 0.943 ($p < .001$). This finding highlighted the overall pool of tax revenues, including corporate, personal, and sales taxes, is a crucial determinant of CCC funding. Dougherty and Natow (2015) emphasized personal income tax and corporate tax fluctuations directly affect public higher education funding, aligning with this study's findings. However, Hovey (2012) did not establish a significant relationship between total tax income and CCC funding. My research, however, offered convincing proof that total state tax income is the main source of community college funding; therefore, future budget planning should consider overall tax revenue trends instead of focusing just on specific tax sources.

## Theoretical and Practical Insights from the Findings

The findings of this study highlight the complex and interdependent relationships between economic indicators, fiscal policies, and funding levels for CCCs. By integrating resource dependence theory (RDT) as the conceptual framework for this research, I was able to analyze how external financial conditions influence the strategic financial planning of community colleges. This study's findings reinforced the central argument of RDT, which posits that organizations rely on external financial resources to achieve their objectives and must continuously adapt to changes in their external environment (Hillman et al., 2009). My results indicated community college funding levels were significantly influenced by state tax collections, especially personal income tax, sales tax, and total tax income, thereby supporting the idea that funding stability is related to more general economic and fiscal conditions.

Emphasizing the importance of financial flexibility and evidence-based strategy planning in response to economic volatility, this study's pragmatic approach fit the findings. Pragmatism motivates organizations to always assess and improve their budgeting procedures to improve their effectiveness and sustainability (Dennes, 1940). The findings revealed GDP growth in dollar terms and CPI in dollar terms are major predictors of CCC funding, meaning macroeconomic trends should be included into financial decisions instead of depending merely on temporary policy changes or reactive budgeting initiatives.

Additionally, my findings challenged previous research by Hovey (2012), which found no significant relationships between GDP, inflation, corporate tax, and CCC funding levels. By refining the measurement approach, using GDP in absolute dollar amounts instead of percentage growth rate and CPI instead of the annual inflation rate, I uncovered relationships that were previously overlooked.

These methodological developments advocate for more accurate and relevant financial forecasting models in higher education, therefore supporting the pragmatic basis of this work.

Furthermore, this study highlighted the implications of these findings for DEI initiatives within CCCs. Given that budget stability directly influences the availability and sustainability of student support programs, DEI offices, and other institutional resources aimed at underrepresented populations, community college leaders must recognize the broader economic forces that shape funding decisions. Without proactive financial planning informed by these findings, DEI programs may be at risk during economic downturns or tax revenue shortfalls.

This study contributed to literature by reinforcing RDT's premise that community colleges depend on external funding sources and must adapt strategically to economic realities. It also emphasized the need of pragmatism in forming responsive and efficient financial planning and supported a data-driven approach to budget forecasting that fits financing models with institutional goals and student achievement. These insights also provided a foundation for policy making and other suggestions, directing community college leaders toward fiscally responsible, fair, and sustainable financial planning models supporting long-term institutional stability. Under the current policy environment, community colleges should negotiate changing financial priorities from the federal government, including a growing focus on workforce development, financial aid expansion, and performance-based funding models. Furthermore, the changing perspective on DEI policies presented fresh difficulties because some states impose limits that might influence financing for programs aiming at student support. Community colleges must create adjustable financial plans that fit evolving federal and state rules to guarantee institutional stability and fair access to education while ensuring institutional integrity.

## Implications

The results of this study provided new perspectives on how financial and political elements influence funding levels of California Community Collections. These discoveries can majorly influence theories in higher education management and leadership styles. This research highlighted the importance of integrating macroeconomic trends into budget forecasting, financial planning, and policy making, by showing the relationships between the variables. This study emphasized the need to include macroeconomic developments in budget forecasting and funding levels. The results of my analysis, moreover, challenge earlier findings on funding factors and highlight the need for more accurate financial modeling to support sustainable funding plans for community colleges. In addition to macroeconomic indicators, the role of administrative services in financial planning is essential for maintaining fiscal stability. Effective administrative control enables universities to maximize budget projections, match resource allocation with institutional priorities, and keep compliance with state and federal financial policies (Barr & McClellan, 2017). Furthermore, performance-based budgets and integrated planning call for administrative coordination to make sure funding decisions support long-term institutional goals. Maintaining funding models and institutional stability will depend on increasing administrative capacity in budget management as community colleges remain under financial stress.

Beyond financial efficiency, the findings also have essential implications for DEI initiatives. Because community colleges serve a diverse student body, including many low-income, first-generation, and underrepresented students, stable funding is essential for sustaining DEI programs. The study suggested policymakers and institutional leaders must prioritize budget forecasting methods that protect these critical student support services from economic downturns and tax revenue fluctuations. Without strategic financial planning, DEI programs may face funding instability, undermining efforts to promote equitable access and student success. These revelations also offered a basis for other recommendations and policy decisions, guiding community college leaders toward fair, sustainable, and fiscally responsible financial planning models endorsing long-term institutional stability. Recent federal policy changes, especially President Trump's fresh attempts to eliminate the U.S. Department of Education, create uncertainty about long-term funding sources for community colleges, though. Greater reliance on state-level budget allocations, which may vary depending on local economic conditions, may result from the suggested decentralization of federal education supervision. Furthermore, affecting funding for student support services, especially those meant to close equity gaps in higher education, are policy changes aimed at eradicating DEI programs. Community colleges must change their financial strategies as these federal programs develop to guarantee sustainable funding in changing legislative and policy environment.

This section explored the implications of the study's findings in four key areas: policy, practice, theory, and future research. The next section discusses how these findings can inform state funding policies and fiscal strategies to enhance financial stability in CCCs.

## Implications for Policy

### *Aligning Funding Models with Economic Growth Trends*

The findings demonstrated a clear, statistically significant relationship between GDP in dollar terms and CCC funding levels, suggesting funding allocation is mostly driven by general economic development. This implies, rather than depending just on fixed allocations or driven by politics budget decisions, state officials should create funding models that are more closely linked to economic performance. Policymakers can guarantee community colleges receive consistent and predictable funding that reflects real economic realities by using dynamic funding formulae that change allocations depending on measures of development, like total GDP in dollar amounts and CPI (Barr & McClellan, 2017). For educational leaders implementing more data-driven financial planning methodologies, this includes macroeconomic indicators into institutional budget forecasting and supporting policies ensuring financing stability in times of economic crisis or recession (Hovey, 2012).

### *Revenue Diversification Strategies*

Following overall tax revenues and sales tax income, personal income tax revenue had the strongest relationship with CCC financing according to this study. Given state budgets vary depending on economic situations, this strong reliance on state tax revenue, especially personal income tax, creates a risk in CCC funding. Policymakers could consider widening revenue diversification techniques include establishing special funding reserves, raising local funding contributions, or investigating other funding sources including public-private partnerships (Zumeta et al., 2021) to help to lower this risk. The reserves act as financial buffers, ensuring funding remains stable during periods of economic downturn or state budget cuts, and raising local contributions involves increasing local property tax allocations, securing municipal grants, or leveraging district-level appropriations to reduce reliance on volatile state tax revenues. Moreover, public–private partnerships (PPPs) create opportunities for community colleges to collaborate with businesses, nonprofit organizations, and philanthropic entities to secure funding for infrastructure, workforce training programs, and student support services. Long-term financial sustainability planning by community college administrators will help to guarantee that institutions are not unduly reliant on erratic state tax collections. In this sense, strategic leadership calls for supporting financing policies that provide community colleges with more financial autonomy so they may create other income sources while keeping student affordability.

### Reforming the Budget Allocation Process to Protect DEI Initiatives

Given CCCs serve a disproportionately large number of low-income, first-generation, and minority students (California Community College Chancellor's Office [CCCCO], 2023), the study emphasized the vital need of financial stability in sustaining DEI programs. DEI programs are typically among the first to see financial cuts when state revenues drop, therefore restricting access to vital support services such as Extended Opportunity Program and Services (EOPS), three original programs founded under Title IV of the Higher Education Act known as (TRIO programs), and disability resource centers (Reckhow et al., 2021). Policymakers should combine equity-oriented financing sources that shield DEI programs from discretionary budget cuts so that budgetary restrictions do not unfairly affect students depending on these services. However, the recent federal push to reduce or eliminate DEI initiatives, particularly under President Trump's proposed restrictions (Faegre Drinker, 2025), poses additional challenges for sustaining these programs. With policy shifts aiming to cut funding for DEI efforts at the federal level, community colleges must explore alternative financial strategies, such as private grants, institutional endowments, and local partnerships, to protect these critical student support services. This policy shift directly affects educational leadership because institutional leaders must actively support financial advocacy to show the long-term advantages of DEI programs in terms of student retention, achievement, and workforce development (Brooks, 2019).

### *Implementing More Accurate Budget Forecasting Models*

The study's findings contradicted other studies (Hovey, 2012) by demonstrating that GDP in dollar values and CPI in absolute values are stronger predictors of CCC financing than GDP growth rate and yearly inflation rate. This finding implied institutional leaders and legislators should use more accurate budget projections that give absolute economic indicators top priority over percentage-based swings in deciding on funding. Conventional budget models can overlook long-term economic cycles, which results in reactive rather than proactive funding plans (Goldstein, 2009). Advanced financial modeling tools like multiyear budget planning and trend-based forecasting help legislators give community colleges more consistent and predictable financing. K–12 school districts and private colleges have increasingly adopted multiyear budget forecasting models that incorporate enrollment projections, demographic trends, and external economic conditions to ensure financial stability. To control income fluctuations, many private colleges also apply tuition elasticity models and endowment management techniques. Similar predictive analytics will help community colleges better control their financial planning and increase funding dependability. Investing in financial data analytics and predictive modeling technologies can help college leaders achieve long-term budgetary sustainability and enhance institutional planning.

### *Enhancing Financial Autonomy for Community Colleges*

This study found corporate tax revenue plays a significant role in CCC funding levels, but corporate tax income is also highly volatile, influenced by market cycles and policy changes. Given these fluctuations, community colleges should have greater financial autonomy to manage their own resources and reduce dependency on unstable state funding (Dougherty & Natow, 2015). Policymakers should explore funding models that grant CCCs more control over local tax revenues, allow greater flexibility in tuition structures, and incentivize revenue-generating partnerships with industry and community organizations. This shift would enable CCC leaders to implement financial strategies that align with institutional goals, rather than being entirely subject to state-level budgetary constraints. In this change, strong financial leadership will be crucial because college presidents and financial officials must negotiate challenging financial environments and make sure that more autonomy does not undermine accessibility and affordability for students.

Through addressing these policy ramifications, state leaders and educational managers can improve institutional resilience, funding equity, and financial sustainability of the California Community College system. Beyond legislative changes, though, these results have major pragmatic ramifications for professors, decision makers, and community college administrators. The next section explores the implications of these findings for educational practice, including financial planning strategies, institutional leadership, and resource allocation models.

## Implications for Practice

The following subsections highlight key areas where these findings can inform institutional practice and strategic planning in community colleges.

### *Strengthening Data-Driven Financial Decision Making*

The findings of this study demonstrated that GDP in dollar amounts and total tax revenue are strong predictors of community college funding, indicating financial decision making should be rooted in macroeconomic and fiscal trend analysis. By including advanced economic forecasting into budget planning procedures, institutional leaders could improve their ability for data-driven decision making (Goldstein, 2009). Investing in financial analytics tools and staff training can help to guarantee administrators, deans, and budget officials have the required ability to understand economic data and project funding trends. Instead of waiting for changes in the budget to guide institutional objectives, community college officials must be proactive. Embedding economic analysis into financial planning would help community colleges achieve evidence-based budgeting, therefore enabling more strategic deployment of resources along with institutional goals and student success priorities.

### *Including Long-Term Budget Forecasting into Architectural Design*

The results of the study show funding levels for community colleges are strongly connected with personal income tax revenue and sales tax income. This connection emphasizes the need of including institutional strategy planning with long-term budget projections. Community colleges should shift from short-term, reactive budgeting to multiyear financial planning models that anticipate economic cycles and revenue fluctuations (Barr & McClellan, 2017). Institutional leaders should collaborate with financial analysts and policymakers to develop budget contingency plans that protect essential academic and student support services during economic downturns. By implementing rolling budget forecasts that project financial conditions over multiple years, community colleges can create more stable, forward-thinking financial strategies that mitigate risks associated with volatile state revenues.

### *Ensuring Financial Stability for DEI and Student Support Programs*

The results of this study highlighted that funding instability, driven by fluctuations in state tax revenue, can place DEI initiatives and student support programs at risk. Because many equity-focused programs (e.g., EOPS, TRIO, disability services) depend on discretionary funding, colleges must adopt financial strategies that ensure long-term sustainability for these critical student success initiatives (CCCCO, 2023). Institutional leaders should prioritize multiyear funding commitments for DEI programs, explore alternative funding sources (e.g., grants, donor contributions), and embed equity-focused budgeting into institutional financial policies. Leadership in this area requires a strong advocacy role, ensuring financial decisions reflect the institution's commitment to diversity, equity, and inclusion, even during periods of economic uncertainty (Brooks, 2019). By including college presidents, vice president of financial management, and board trustees in long-term financial plans, DEI projects' sustainable funding is mostly guaranteed. To keep these initiatives running, academic deans and student affairs managers must cooperate to find other funding sources (e.g., focused grants, donor contributions). By offering data-driven justifications for ongoing DEI project investment and proving their influence on student retention, achievement, and workforce outcomes, institutional research offices can also help.

### Building Financial Resilience Through Revenue Diversification

Because CCC funding heavily depends on state tax revenue, especially personal income tax, institutional leaders should adopt financial diversification strategies to reduce reliance on unpredictable state funding. Community colleges can strengthen partnerships with local industries, expand revenue-generating workforce development programs, and pursue alternative funding through grants and philanthropic contributions (Zumeta et al., 2021). Establishing endowment funds could also provide financial security during economic downturns. Strong financial leadership requires a strategic mindset, ensuring institutions are not overly dependent on a single funding stream but instead leverage multiple revenue sources to sustain operations and student success initiatives.

### Enhancing Collaboration Between Financial Officers and Academic Leadership

Study findings highlighted the need for institutional budgeting to consider economic trends and fiscal policy changes. These findings underscored the importance of collaboration between financial officers, academic deans, and student services administrators. Traditionally, financial planning and academic planning operate in separate silos, leading to misalignment between budget allocations and institutional priorities (Goldberg & Prottas, 2017). By using an integrated financial planning methodology, community colleges can close this disparity and guarantee academic leaders and financial managers cooperate to match programmatic needs and student success goals with budget allocation. Integrated financial planning methodology refers to a strategic approach that aligns budgeting processes with institutional priorities by fostering collaboration between financial officers and academic leadership. This method involves multiyear financial forecasting, data-driven budget allocation, and cross-departmental coordination to ensure fiscal sustainability while meeting student success objectives. By integrating academic planning with financial decision making, institutions can create more adaptable budget frameworks that respond to enrollment trends, state funding changes, and economic fluctuations. This collaborative approach strengthens institutional resilience, enhances budget transparency, and ensures that funding priorities align with financial realities and educational objectives.

Community colleges may build more sustainable financial systems supporting institutional goals and student success by using these data-driven financial planning tools, income diversification models, and equity-centered budgeting approaches. Beyond useful applications, nevertheless, these results have more general ramifications for educational finance theory, especially with relation to resource reliance, strategic financial planning, and the changing function of economic forecasting in higher education budgeting. The theoretical ramifications of these results are discussed in the next part, which also provides understanding of how this study supports the continuous conversation in strategic planning and financial management in higher education.

## Implications for Theory

Beyond practical applications, the findings of this study also carry significant implications for theory. These implications help deepen the understanding of how external economic variables influence institutional funding behavior and strategic planning frameworks in higher education.

### *Refining RDT in the Context of Higher Education Finance.*

The findings of this study support and extend RDT by demonstrating community colleges are highly dependent on external fiscal conditions, particularly state tax revenues and macroeconomic trends, in determining their funding levels (Hillman et al., 2009). My findings implied community college funding has been influenced by personal income, tax income, sales tax income, and corporate tax income, underlining the need of institutions to consistently change to meet outside financial constraints. This study also implied not all outside economic variables have the same degree of impact. Although total tax income and GDP in dollar amounts strongly correlated with funding levels, other variables (e.g. the unemployment rate, GDP growth rate) did not demonstrate significant relationships. These findings suggested RDT should be refined in the context of higher education finance by distinguishing between financial dependencies that have a direct impact on funding allocations versus those that are more peripheral. Future theoretical frameworks should consider differentiating between short-term economic fluctuations (e.g., unemployment rates) and structural financial dependencies (e.g., tax revenue streams) when applying RDT to public higher education institutions.

### Expanding Pragmatism as a Framework for Strategic Financial Decision Making

This study also extends pragmatism as a theoretical framework by emphasizing the importance of data-driven, adaptive financial planning in response to external economic conditions (Dennes, 1940). Pragmatism suggests that leaders must take action based on real-world conditions, continuously testing and refining strategies to improve efficiency and effectiveness. The study findings demonstrated community college financial planning must shift toward a pragmatic, evidence-based approach that integrates macroeconomic indicators into budget forecasting. Traditional budgeting models in higher education often rely on historical allocations and political negotiations rather than economic realities (Goldstein, 2019). Nonetheless, my research revealed high correlations between CCC funding and macroeconomic data (e.g., CPI and GDP in dollar values) point to the need of predictive financial models based on pragmatic ideas. The results showed educational finance theory should change to include economic forecasting as a primary element of strategic financial planning, therefore guaranteeing institutions stay flexible and financially strong.

### Bridging Economic Forecasting and Higher Education Budget Theory

The results of this study suggested traditional higher education budget models do not adequately account for economic forecasting as a central factor in funding allocations. My findings showed state tax revenues and economic growth patterns serve as stronger predictors of CCC funding levels than previously acknowledged, calling for a more dynamic budgetary framework that integrates financial forecasting tools (Zumeta et al., 2021). Although conventional models of higher education finance rely on funding formulas, tuition policies, and political impacts, they sometimes ignore the forecasting power of macroeconomic trends (Barr & McClellan, 2017). The findings of this research implied future theoretical models should include into budgetary decision-making systems economic forecasting systems including GDP in absolute values, CPI, and tax income trends. Higher education financial theories can change in response to this to better handle the intricate interaction between institutional financing stability and economic cycles.

### *Integrating Equity Considerations into Higher Education Finance Theories*

Emphasizing the significance of including DEI ideas into higher education finance theory, this study also showed the important intersection between financial stability and fairness in community institutions. Because community colleges serve a disproportionately large number of low-income, first-generation, and underrepresented students, funding fluctuations directly impact the availability of essential student support services (CCCCO, 2023). Existing financial theories often focus on efficiency and sustainability but do not fully incorporate equity-based financial planning. My results implied future theoretical models should consider the function of financial stability in safeguarding DEI efforts, therefore guaranteeing economic downturns do not disproportionately affect students from disadvantaged backgrounds. Higher education finance theories should include equity-centered budgeting models to help academics and legislators create more comprehensive models that successfully achieve a mix between fair resource allocation and financial sustainability. For example, finance theories that incorporate equity considerations include vertical equity theory, which argues that institutions with greater student need should receive higher funding allocations, and need-based resource allocation models, which prioritize funding based on student demographics and institutional support requirements to reduce financial disparities

These theoretical consequences highlight the requirement of a more complex knowledge of financial interdependence, the part of economic forecasting in budgeting, and the incorporation of DEI ideas into higher education finance models. Although this study contributed to these discussions, it also raised several important questions for future research. The next section explores directions for future research, including areas where additional empirical studies can refine and expand upon the findings presented here.

**Implications for Future Research**

To guide future exploration, this section outlines specific areas where continued research can expand the understanding of funding mechanisms and their long-term effects on CCCs.

***Examining the Long-Term Stability of Tax Revenue as a Funding Source***

This study found personal income tax revenue had the strongest correlation with community college funding, followed by total tax income and sales tax revenue. Although these findings confirmed the reliance of CCCs on state tax revenue, future research should explore the long-term sustainability of these funding sources, particularly in the context of economic downturns, demographic shifts, and changes in tax policy (Zumeta et al., 2021). Given that personal income tax is highly susceptible to labor market conditions, future studies could analyze how economic recessions and workforce trends impact the stability of CCC funding. Researchers should also investigate alternative revenue models that could supplement state tax income, such as local funding contributions, industry partnerships, and performance-based funding structures. Such research would help policymakers and institutional leaders develop more resilient financial strategies that ensure long-term funding stability for CCCs.

***Exploring the Relationship Between Funding Stability and Student Outcomes***

Although this study focused on the financial factors of CCC funding, future studies should also look at how variations in funding affect student success indicators like retention, completion rates, and workforce outcomes. Existing research has suggested financial instability in community colleges can negatively impact student support services, faculty hiring, and program availability, which in turn affects student performance (Goldrick-Rab, 2016). Future studies should use longitudinal data to assess whether changes in funding levels, particularly during economic downturns, correlate with disparities in student achievement and equity gaps. Echoing the requirement of consistent and predictable budget models in higher education, this line of study would offer important new insights on how financing decisions directly affect institutional effectiveness and student outcomes.

### *Investigating the Role of Financial Autonomy in Community College Funding Models*

Given that this study highlighted the strong influence of state tax revenue on CCC funding, another area for future research is the extent to which increased financial autonomy for community colleges could improve funding stability. Although CCCs primarily rely on state appropriations, some states have experimented with local funding mechanisms, expanded grant programs, and partnerships with private industry to create more diversified revenue streams (Dougherty & Natow, 2015). Future studies should investigate whether giving CCCs more financial resource control (e.g., permitting local sales tax allocations or performance-based incentives) could improve their capacity to sustain steady funding despite changes in state finances. Comparative case studies across different states could help identify best practices for balancing financial autonomy with affordability and accessibility in public higher education.

### *Assessing the Efficacy of Budget Forecasting Models in Higher Education*

The findings of this study demonstrated GDP in dollar amounts and CPI are strong predictors of CCC funding, whereas GDP growth rate and unemployment rates are not. These results suggested economic forecasting models used in higher education finance should prioritize absolute economic indicators over percentage-based fluctuations. Future research should evaluate the effectiveness of different budget forecasting models in accurately predicting community college funding trends. Specifically, researchers could compare traditional linear models with more advanced forecasting techniques (e.g., machine learning algorithms, scenario-based projections) to determine which models provide the most accurate funding predictions. This research would be valuable for policymakers and institutional leaders seeking to implement data-driven financial planning strategies that account for economic volatility.

*Exploring the Intersection of Financial Stability and DEI in Community Colleges*

The study findings indicated funding instability places equity-focused programs (e.g., DEI initiatives, EOPS, workforce development programs) at risk. Future research should investigate how financial fluctuations impact the availability and effectiveness of programs designed to support low-income, first-generation, and underrepresented students. By analyzing funding allocation patterns, researchers can identify whether budget reductions disproportionately affect student populations who rely on institutional support services. Additionally, future studies should examine the role of targeted funding policies (e.g., equity-based budgeting, need-based grant allocations) in protecting DEI programs from financial cuts. This research would provide valuable insights into how funding models can be structured to ensure financial sustainability while advancing equity and inclusion in community colleges.

These future directions underline the need of more investigation of financing stability, student performance, financial autonomy, and equity in community institutions. Although this study offers important new perspectives on the fiscal and economic factors influencing CCC funding, more study is required to improve financial forecasting models, assess different income sources, and create policies guaranteeing financial resilience in higher education. Although Hovey (2012) found no significant relationship between GDP growth rates and community college funding levels, this study demonstrated GDP in dollar values and CPI have stronger predictive power. These findings challenge prior assumptions by emphasizing the need for absolute economic indicators over percentage-based fluctuations in forecasting models. Understanding these dynamics will be essential for policymakers, institutional leaders, and researchers committed to strengthening the long-term financial sustainability of community colleges. Building on these implications, the following part offers suggestions for financial planners, community college leaders, and legislators to implement these conclusions in pragmatic ways that improve fiscal stability, resource allocation, and equity-driven budgeting practices.

## Recommendations

To strengthen funding stability, this section offers actionable recommendations that align institutional finance practices with the study's key findings.

### Develop a Data-Driven, Dynamic Funding Model

This study revealed CCC funding levels significantly correlate with GDP in dollar terms and total tax revenues. But present funding structures do not fairly represent real-time economic situations, which results in erratic funding changes. Macroeconomic data like GDP, CPI, and state tax collections should be included into a data-driven, dynamic funding model adopted by state legislators into budget allocation decisions (Barr & McClellan, 2017). Predictive modeling and real-time financial data help funding decisions to become more consistent, sensitive, and in line with actual economic performance. Leaders in community colleges should support budget ideas that change allocation depending on economic data instead of depending merely on set legislative funding. This approach would guarantee that financial reserves are created during economic downturns to preserve stability and that financing increases match economic development.

### Strengthen Revenue Diversification and Financial Autonomy

Given the study's findings that CCC funding is heavily dependent on state tax revenues, particularly personal income tax ($R^2 = 0.889$, $p < .001$), community colleges must reduce their vulnerability to revenue volatility. Institutions should explore alternative revenue sources such as industry partnerships, workforce development grants, endowment funds, and local funding initiatives (Dougherty & Natow, 2015). Policymakers should also provide community colleges with more autonomy, allowing them to manage local revenue streams, set differential tuition structures for specialized programs, and develop innovative funding mechanisms. Increasing financial independence can enable CCC officials to make long-term financial decisions consistent with institutional interests and reduce the risks connected with changes in state tax income.

## Institutionalize Equity-Centered Budgeting for DEI Programs

The study's findings highlighted financial instability disproportionately affects DEI programs and student support services. As budget cuts often target discretionary funding, equity-focused initiatives (e.g., EOPS, TRIO, disability resource centers) and basic needs programs face persistent financial threats (CCCCO, 2023). Institutions should institutionalize equity-centered budgeting by embedding long-term financial commitments for DEI programs into baseline budgets rather than treating them as expendable line items. Policymakers should also establish protected funding pools for equity-focused programs to ensure financial sustainability, even during economic downturns. Additionally, educational leaders must advocate for DEI-conscious financial policies that recognize the long-term benefits of these programs in promoting student success and economic mobility (Brooks, 2019).

## Expand Predictive Budgeting and Financial Forecasting Training

This study revealed reactive financial decision-making results from standard budgeting methods not considering predictive economic data. Programs for thorough financial forecasting training for budget officials, academic leaders, and administrators should be instituted by community colleges (Hovey, 2012). Leadership training courses should incorporate professional development in predictive budgeting models, economic forecasting, and data analytics into order to improve financial literacy and planning capacity. Advanced financial modeling methods that let for trend-based funding strategies, economic scenario planning, and multiyear revenue estimates should also be adopted by institutions. This proactive approach would help institutional leaders anticipate financial challenges, optimize resource allocation, and ensure long-term fiscal sustainability (Goldstein, 2009).

## Enhance Collaboration Between Educational Leaders and Policymakers

The study drew attention to a discrepancy between institutional financial reality and state-level budget strategies. Leaders of community colleges must be more involved in financial policy advocacy if they are to close this disparity. Stronger partnerships between institutional leaders, state legislators, and financial policymakers can ensure budget decisions reflect the actual financial needs of community colleges rather than being dictated solely by political cycles (McKeown-Moak & Mullin, 2014). Establishing policy advisory committees comprising community college presidents, financial managers, and student representatives may help to increase funding distribution clarity and enhance cooperative decision making. Regular communication between government agencies and higher education institutions helps legislators create funding plans that complement institutional missions, workforce development goals, and student success results.

These recommendations provide a strategic roadmap for improving the financial sustainability, equity, and resilience of CCCs. Following these suggestions would support long-term financial planning, protect DEI projects, and aid to stabilize funds. Providing the study's final result, the last part lists important conclusions and their wider ramifications for educational leadership, policy, and future research. Community college leaders must take proactive steps to ensure financial resilience in the face of fluctuating state revenues. This entails supporting steady, long-term funding models that give equity top priority, including multiyear financial projections into institutional planning, and building closer ties with legislators to guarantee specific financing for projects aimed at student success. Following these recommendations will help institutions safeguard DEI programs and financial aid resources and increase their capacity to offer all students an accessible, top-notch education. These initiatives will guide educators in maintaining California's community colleges through economic uncertainty and ensuring financial decisions reflect inclusiveness, workforce development, and academic excellence goals.

## Summary of the Dissertation

This study explored the intricate relationship between economic indicators, fiscal policies, and funding levels for CCCs. This research's central focus was the financial instability CCCs face due to shifting economic conditions, unpredictable tax revenues, and inadequate forecasting methods. Without reliable funding, these institutions may struggle to provide affordable, accessible education, especially for students from historically underrepresented backgrounds. This study aimed to address this challenge by identifying the key economic and fiscal factors that influence CCC funding and offering data-driven recommendations for financial sustainability.

Through a quantitative analysis, I found GDP in dollar amounts, CPI, and state tax revenues (e.g., personal income tax, sales tax, total tax income) have significant predictive power over CCC funding levels, whereas other economic indicators (e.g., unemployment rates, GDP growth rate) showed no significant relationship. These results underline the need of more dynamic, data-driven budgeting models reflecting real-time economic situations and challenge earlier presumptions about the financial drivers of community college funding. Particularly in guaranteeing the longevity of DEI programs, worker training efforts, and student support services, the results of the study further emphasize the vital link between financial stability and student achievement.

Based on these findings, I proposed five key recommendations: (a) develop a data-driven, dynamic funding model that ties CCC funding to macroeconomic trends; (b) strengthen revenue diversification and financial autonomy to reduce reliance on volatile state tax revenues; (c) institutionalize equity-centered budgeting to protect DEI programs from financial instability; (d) expand predictive budgeting and financial forecasting training for institutional leaders; and (e) enhance collaboration between policymakers and community college administrators to ensure funding policies align with institutional needs. In higher education, these recommendations offer a strategic framework for enhancing policy creation, budget forecasting, and financial planning.

Envision once again the bright-eyed student stepping onto a college campus for the first time, carrying the hopes and dreams of their family. The reality of financial instability in community colleges means students like this, particularly first-generation, low-income, and underrepresented students, face barriers not just in accessing education, but in persisting through it. Funding decisions made at the state, institutional, and policy levels have real consequences on whether these students can enroll in the classes they need, receive essential support services, and ultimately, walk across the stage at graduation. Every dollar in a community college budget represents not just an institutional expense, but an opportunity, a chance for students to break generational cycles, enter the workforce, and contribute meaningfully to society. This study emphasized the critical necessity of sustainable, comprehensive, equity-driven financial planning in community colleges so that every student who steps onto campus has the tools they need to achieve, independent of financial situation.